ORIGINAL PAPER

# Extremal cases of distortion risk measures with partial information

Mengshuo Zhao[1,2] · Narayanaswamy Balakrishnan[3] · Chuancun Yin[1] · Hui Shao[4]

Received: 27 January 2026 / Accepted: 19 June 2026

## Abstract

This paper investigates the impact of distributional uncertainty on key risk measures under the partial knowledge of underlying distributions characterized by their first two moments and shape information (specifically symmetry and/or unimodality). We first employ probability inequalities to establish the theoretical best- and worst-case bounds on Value-at-Risk, reflecting the most extreme tail risk achievable within the moment and shape constraints, and then we extend this worst-case/best-case analysis to a broad class of distortion risk measures by the modified Schwarz inequality, deriving their corresponding robust bounds under the same partial information setting concerning moments and distribution shapes of the underlying distributions. In addition, we give a clear characterization of the distributions that attain the best- and worst-case scenarios. The proposed approach provides a unified framework for extremal problems of distortion risk measures.

✉ Chuancun Yin
ccyin@qfnu.edu.cn

✉ Hui Shao
mathshao@gmail.com

Mengshuo Zhao
mengshuozhao@qfnu.edu.cn

Narayanaswamy Balakrishnan
bala@mcmaster.ca

1 School of Statistics and Data Science, Qufu Normal University, Qufu 273165, Shandong, China

2 School of Insurance, Central University of Finance and Economics, Beijing 100081, China

3 Department of Mathematics and Statistics, McMaster University, Hamilton, ON, Canada

4 International Business School, Zhejiang University, Haining 314400, China

# 1 Introduction

Risk measures serve as fundamental tools in quantitative risk management and are widely employed in insurance and finance. Value-at-Risk (VaR) and Tail Value-at-Risk (TVaR) are two of the most widely accepted risk measures in financial and insurance industries (see, e.g., Denuit et al. (2005)). These two risk measures are unified in a more general two-parameter family of risk measures, called the Range-Value-at-Risk (RVaR). The family of RVaR was introduced by Cont et al. (2010) in the context of robustness properties of risk measures. A more general risk measure, called the distortion risk measure (DRM) developed from the research on premium principles by Denneberg (1990), has been widely used in behavioral economics and risk management as an alternative to expected utility. By choosing appropriate distortion functions, the distortion risk measure offers a unified framework for quantile-based risk measures, and it includes a range of prominent risk measures such as VaR, TVaR and RVaR as special cases.

To derive the quantile-based risk measures, one often requires an exact form of the distribution of the specific risk, which is usually not feasible in practice. A pragmatic approach involves working with partial information-such as the mean, variance, and shape characteristics-rather than the full distribution, thereby giving rise to the problem of distributional uncertainty. For further background on risk bounds and model uncertainty in quantitative finance and insurance, we refer to Rüschendorf et al. (2024). This prompts researchers to develop risk measures for extreme cases that require only partial information. The problem of finding sharp bounds for risk measures under model uncertainty has received much attention in the literature. Early contributions include El Ghaoui et al. (2003), who derived a closed-form solution for worst-case VaR under known mean and variance, and Chen et al. (2011), who obtained closed-form bounds for worst-case TVaR under the same constraints. Li (2018) later extended these results to a broad class of law-invariant coherent risk measures. Subsequent research has expanded this framework to incorporate higher-order moment information. For instance, Liu et al. (2020) derived worst-case values for law-invariant convex risk functionals with knowledge of the mean and higher-order moments. More recently, Cai et al. (2025) generalized earlier results to arbitrary distortion risk measures, while Cai et al. (2024) studied worst-case risk measures for stop-loss and limited loss variables with applications in robust reinsurance. Some recent work by Pesenti et al. (2024) considered optimizing distortion riskmetrics with distributional uncertainty, and Han and Liu (2025) investigated the robust models for $\Lambda$-quantiles with partial information regarding the loss distribution.

Another significant research direction incorporates shape information to derive tighter and more realistic risk bounds. By introducing realistic distributional characteristics such as unimodality and symmetry, researchers have achieved substantial improvements in bounding accuracy. Li (2018) established bounds for RVaR under symmetry and unimodality assumptions, while Zhu and Shao (2018) generalized these findings to distortion risk measures with symmetry constraint, considering both worst-case and best-case bounds. Recently, Bernard, Kazzi, and Vanduffel have done a series of work on deriving RVaR bounds under specific shape constraints, moving from unimodality with moments (Bernard et al. 2020), to unimodality with sym-metry (Bernard et al. 2022), and ultimately to RVaR and VaR bounds for unimodal T-symmetric distributions (Bernard et al. 2025a, b). Bernard et al. (2024) derived quasi explicit best- and worst-case values of a large class of distortion risk measures when the underlying loss distribution has a given mean and variance and lies within a $\sqrt{\varepsilon}$-Wasserstein ball around a given reference distribution, which generalizes the results of Li (2018) and Zhu and Shao (2018) corresponding to a Wasserstein tolerance of $\varepsilon = +\infty$.

Further contributions in this vein include (Shao and Zhang 2023, 2024), who used the first two moments (or certain higher-order absolute center moments) and symmetry assumptions to derive closed-form extremal distortion risk measures and characterized the corresponding extremal distributions via distortion function envelopes. Zuo and Yin (2025) studied worst-case distortion riskmetrics and weighted entropy under mean-variance information. Han and Liu (2025) established sharp upper and lower bounds for distortion risk metrics under distributional uncertainty. The uncertainty sets are characterized by four key features of the underlying distributions: mean, variance, unimodality, and Wasserstein distance to a reference distribution.

Most of the existing works only consider the worst-cases of the special risk measures such as VaR, TVaR, RVaR and the coherent risk measures, while the best-case analysis and the treatment of general distortion risk measures have received less attention. In addition, under the combined symmetry and unimodality constraint, results are available only for RVaR and its special cases, but not for general DRMs. Motivated by the above-mentioned work, this paper aims to study both the worst-case and best-case estimates of any distortion risk measures based on partial information about the underlying distributions. Firstly, we attempt to directly calculate the worst- and best-case estimates of VaR under certain assumptions concerning partial moment and shape constraints. Then, we want to apply these findings to extreme estimates of distortion risk measures by solving the optimization problems based on four types of assumptions on the underlying distributions. As a result, we obtain analytical worst- and best-case bounds that provide a unified framework for the existing bounds and thus expand the existing results on risk measures under various shape constraints. The framework covers the previously known bounds for VaR, TVaR, and RVaR under various shape constraints as special cases, and is applicable to any distortion risk measure by inserting the corresponding distortion function. The problems can be very challenging and require various novel methodologies for different model settings. Most of the literature on this topic uses the standard duality method in optimization theory to obtain the results under the convexity assumption of distortion functions (see, e.g., El Ghaoui et al. 2003; Popescu 2005; Li 2018) or via a unified method combining convex order and the notion of joint mixability (see Li (2018)).

The main contribution of this paper is three-fold:

• First, we use a set of elegant probabilistic inequalities to systematically derive VaR and $\mathrm{VaR}^+$ bounds-along with their attaining distributions-under four distributional uncertainty sets which are defined by mean and variance, symmetry, unimodality, unimodal-symmetry of the underlying distributions. Unlike previous studies that often treat worst- and best-case VaR and $\mathrm{VaR}^+$ separately, we provide a unified treatment using elegant probabilistic inequalities. This not only recovers but also improves several classical results (e.g., El Ghaoui et al. 2003; Li 2018; Bernard et al.

2020, 2022) and, importantly, explicitly characterizes the extremal distributions that attain the bounds, a feature missing in some earlier works.

• Second, we use various calculus techniques and the modified Schwarz inequality to derive the extreme-case DRMs and their corresponding distributions for all distributions/symmetric distributions when mean and variance are available.

• Third, we study the extreme-case DRMs with general distortion functions under the unimodality/symmetry and unimodality constraints. The answer to this question is still missing in the literature and results are available only for special cases (e.g., VaR, TVaR, RVaR).

The rest of this paper is structured as follows. Section 2 introduces the necessary notation and formulates the problem. Section 3.1 examines the extreme-case VaR under four types of assumptions on the underlying distributions. The main results—the analytical worst-case and best-case estimates and their corresponding distributions of distortion risk measures under partial moments and shape constraints—are presented in Sect. 3.2. In Sect. 3.3, we calculate the bounds of VaR, TVaR and RVaR using the unified result obtained in the preceding section. Section 3.4 provides a simple numerical illustration. Finally, some concluding remarks are presented in Sect. 4.1.

## 2 Preliminaries

Let $X$ be a non-degenerate random variable with distribution function $F_X$, finite mean $\mu$ and finite positive variance $\sigma^2$. We define the right-continuous generalized inverse of $F_X$ as

$$F_X^{-1+}(p) = \sup\{x : F_X(x) \leq p\},\ 0 \leq p < 1,$$

whereas the left-continuous generalized inverse of $F_X$ is defined as

$$F_X^{-1}(p) = \inf\{x : F_X(x) \geq p\},\ 0 < p \leq 1,$$

with $\inf\{\varnothing\} = +\infty$ and $\sup\{\varnothing\} = -\infty$ by convention. For any real number $x$ and probability level $p$, it is well known that $x \leq F_X^{-1+}(p) \iff P(X < x) = F_X(x-) \leq p$ and $F_X^{-1}(p) \leq x \iff p \leq F_X(x)$. It is easy to see that $F_X^{-1}(p) \leq F_X^{-1+}(p)$, there are at most countably many values of $p \in [0, 1]$ where $F_X^{-1}(p)$ and $F_X^{-1+}(p)$ are different (see, e.g., Dhaene et al. (2012)). Moreover, $F_X^{-1}(p) = F_X^{-1+}(p)$ iff $F_X(x) = p$ for at most one $x$.

The right-continuous Value-at-Risk at level $\alpha \in (0, 1)$ of a risk $X$ is defined as

$$\mathrm{VaR}_\alpha^+[X] = F_X^{-1+}(\alpha),\ 0 \leq \alpha < 1,$$

whereas the left-continuous Value-at-Risk at level $\alpha \in (0, 1)$ of a risk $X$ is defined as

$$\mathrm{VaR}_\alpha[X] \equiv \mathrm{VaR}_\alpha^-[X] = F_X^{-1}(\alpha),\ 0 < \alpha \leq 1.$$

The Tail Value-at-Risk (TVaR) of a risk $X$ is defined as

$$\mathrm{TVaR}_\alpha[X] = \frac{1}{1-\alpha}\int_\alpha^1 F^{-1}(p)dp = \frac{1}{1-\alpha}\int_\alpha^1 F^{-1+}(p)dp,\ 0 < \alpha < 1.$$

The Range Value-at-Risk (RVaR) is defined for $\alpha, \beta \in (0, 1)$ as (see Cont et al. (2010))

$$\mathrm{RVaR}_{\alpha,\beta}[X] = \frac{1}{\beta-\alpha}\int_\alpha^\beta F^{-1}(p)dp = \frac{1}{\beta-\alpha}\int_\alpha^\beta F^{-1+}(p)dp,\ 0 < \alpha < \beta < 1.$$

Clearly, RVaR includes TVaR and VaR as limiting cases.

**Definition 1** (Denuit et al. 2005) Given a random variable $X$ with distribution $F_X$, the distortion risk measure (DRM) of $X$ is defined via the Choquet integral in the form

$$\rho_h[X] = \int_0^\infty h(\bar{F}_X(x))dx + \int_{-\infty}^0 (h(\bar{F}_X(x)) - 1)dx,$$

whenever at least one of the two integrals is finite. The function $h$ refers to a distortion function which is a nondecreasing function on [0, 1] with $h(0) = 0$ and $h(1) = 1$. It is well-known that DRM is a law invariant, positively homogeneous, monotone and comonotonic additive risk measure. If $h$ is concave, then $\rho_h[X]$ is a coherent risk measure.

**Definition 2** ( Boyd and Vandenberghe 2004) For a distortion function $h$, the convex and concave envelopes of $h$ are defined, respectively, by

$$h_* = \sup\{g | g : [0, 1] \to [0, 1] \text{ is convex and } g(p) \leq h(p), p \in [0, 1]\},$$

$$h^* = \inf\{g | g : [0, 1] \to [0, 1] \text{ is concave and } g(p) \geq h(p), p \in [0, 1]\}.$$

Note that $(-h)_* = -h^*$. Moreover, if $h$ is convex, then $h_* = h$; if $h$ is concave, then $h^* = h$.

**Definition 3** ( Khintchine 1938) A random variable $X$ (or its corresponding distribution) is called unimodal if there exists a constant $m \in \mathbb{R}$ (called the mode) such that its distribution function $F_X$ is convex-concave—convex on $(-\infty, m)$ and concave on $(m, +\infty)$. A random variable $X$ is called symmetric if there exists a constant $m \in \mathbb{R}$ (the symmetric center) such that $P(X \leq x) = P(X \geq 2m - x)$ for any $x \in \mathbb{R}$.

For $(\mu, \sigma) \in \mathbb{R} \times \mathbb{R}^+$, we denote by $V(\mu, \sigma)$ the set of random variables with mean $\mu$ and variance $\sigma^2$, and denote by $V_S(\mu, \sigma)$, $V_U(\mu, \sigma)$ and $V_{US}(\mu, \sigma)$ the sets of random variables in $V(\mu, \sigma)$ with symmetric, unimodal, symmetric and unimodal distribution functions, respectively. These shape constraints are motivated by their prevalence in parametric modeling: many standard distributions are symmetric, unimodal, or both (e.g., normal and Student's $t$ are symmetric-unimodal; Gamma and log-normal are unimodal).

For given distortion function $h$, we consider the following optimization problems (worst-case and best-case)

$$\sup_{X\in\mathcal{V}(\mu,\sigma)} \rho_h[X] \text{ and } \inf_{X\in\mathcal{V}(\mu,\sigma)} \rho_h[X],$$

respectively, where $\mathcal{V}(\mu,\sigma)$ denotes one of $V(\mu,\sigma)$, $V_S(\mu,\sigma)$, $V_U(\mu,\sigma)$ and $V_{US}(\mu,\sigma)$. If a random variable $X_*$ satisfies $\sup_{X\in\mathcal{V}(\mu,\sigma)} \rho_h[X] = \rho_h[X_*]$, then we refer to $F_{X_*}$ as a worst-case distribution and $X_*$ as a worst-case rv. Similarly, we refer to $F_{X^*}$ as a best-case distribution if $\inf_{X\in\mathcal{V}(\mu,\sigma)} \rho_h[X] = \rho_h[X^*]$, and $X^*$ is called a best-case rv. The values $\sup_{X\in\mathcal{V}(\mu,\sigma)} \rho_h[X]$ and $\inf_{X\in\mathcal{V}(\mu,\sigma)} \rho_h[X]$ are correspondingly the worst-case and best-case distortion risk measures, respectively.

## 3 VaR bounds

The worst-case and best-case VaR under constraints on the first two moments have been extensively studied. The known results are scattered in different publications; see, e.g., (Ion et al. 2023; Li 2018; Zhu and Shao 2018; Bernard et al. 2020, 2022, 2025a, b; Shao and Zhang 2023, 2024; Pesenti et al. 2024; Cai et al. 2025), and so on. The literature treats the worst-case and best-case estimates separately, and some even have flaws in proof. It should be also noted that (Bernard et al. 2020, 2022, 2025a) derived extreme-case VaR and RVaR under distributional uncertainty sets defined by a fixed mean, a bounded variance, and either unimodality or unimodal symmetry. In this section, we consider the worst-case and best-case VaR risk measures with the first two moments, as well as symmetry and unimodality of the underlying distribution. Our results unify and improve many existing results in the literature. The methods we used are different from the ones in the literature, mainly using the sharp upper bounds on the tails of the distribution of $X$; see Ion et al. (2023) for more such sharp upper bounds.

**Notation.** Throughout this section, we denote by $\mathrm{VaR}^+_\alpha[X] = F_X^{-1+}(\alpha)$ the right-continuous VaR and by $\mathrm{VaR}_\alpha[X] = \mathrm{VaR}^-_\alpha[X] = F_X^{-1}(\alpha)$ the left-continuous VaR, as introduced in Section 2. We write $\mathrm{VaR}^\pm_\alpha[X]$ when a statement holds identically for both versions. Since the extremal distributions constructed below are often discrete or mixed, the two definitions may differ in attainability; whenever this occurs, we treat $\mathrm{VaR}^+_\alpha[X]$ and $\mathrm{VaR}_\alpha[X]$ separately and comment on the distinction.

### 3.1 Case of general distributions

We shall use the Cantelli inequality:

**Lemma 3.1** *(Ion et al. 2023) Let X be a random variable with mean $\mu$ and variance $\sigma^2$. Then, for any $x \geq \mu$, the inequality*

$$P(X \geq x) \leq \frac{\sigma^2}{\sigma^2 + (x-\mu)^2}$$

*holds. The equality is uniquely attained by*

$$P(X = x) = \frac{\sigma^2}{\sigma^2 + (x-\mu)^2},\ P\left(X = \mu - \frac{\sigma^2}{x-\mu}\right) = \frac{(x-\mu)^2}{\sigma^2 + (x-\mu)^2}.$$

The following proposition gives the worst- and best-case estimates for $\mathrm{VaR}_\alpha^+[X]$ and $\mathrm{VaR}_\alpha[X]$.

**Proposition 3.1** *Let X be a random variable with mean $\mu$ and variance $\sigma^2$. Then, the following hold: (i) We have, for $0 < \alpha < 1$,*

$$\sup_{X \in V(\mu,\sigma)} \mathrm{VaR}_\alpha^+[X] = \sup_{X \in V(\mu,\sigma)} \mathrm{VaR}_\alpha[X] = \mu + \sigma\sqrt{\frac{\alpha}{1-\alpha}}, \tag{3.1}$$

*and the first supremum in (3.1) is attained at the two-point distribution*

$$P\left(X_* = \mu + \sigma\sqrt{\frac{\alpha}{1-\alpha}}\right) = 1-\alpha,\ P\left(X_* = \mu - \sigma\sqrt{\frac{1-\alpha}{\alpha}}\right) = \alpha.$$

*(ii) We have, for $0 < \alpha < 1$,*

$$\inf_{X \in V(\mu,\sigma)} \mathrm{VaR}_\alpha[X] = \inf_{X \in V(\mu,\sigma)} \mathrm{VaR}_\alpha^+[X] = \mu - \sigma\sqrt{\frac{1-\alpha}{\alpha}}, \tag{3.2}$$

*and the first infimum in (3.2) is attained at the two-point distribution*

$$P\left(X^* = \mu - \sigma\sqrt{\frac{1-\alpha}{\alpha}}\right) = \alpha,\ P\left(X^* = \mu + \sigma\sqrt{\frac{\alpha}{1-\alpha}}\right) = 1-\alpha.$$

***Proof*** Without loss of generality, we only prove the case when $\mu = 0$ and $\sigma = 1$. It follows from Cantelli's inequality that, for any $x \geq 0$,

$$P(X \leq x) \geq \frac{x^2}{1+x^2}.$$

Letting $\alpha = \frac{x^2}{1+x^2}$, then $x = \sqrt{\frac{\alpha}{1-\alpha}}$. As $F(x) \geq \alpha$, we get $\mathrm{VaR}_\alpha[X] \leq \sqrt{\frac{\alpha}{1-\alpha}}$. Clearly,

$$P(Z=x)=\frac{1}{1+x^2},\ P\left(Z=-\frac{1}{x}\right)=\frac{x^2}{1+x^2},$$

if and only if

$$P\left(Z=\sqrt{\frac{\alpha}{1-\alpha}}\right)=1-\alpha,\ P\left(Z=-\sqrt{\frac{1-\alpha}{\alpha}}\right)=\alpha.$$

In this case, $\mathrm{VaR}_\alpha^+[Z]=\sqrt{\frac{\alpha}{1-\alpha}}$ and $\mathrm{VaR}_\alpha[Z]=-\sqrt{\frac{1-\alpha}{\alpha}}$. Therefore, for the arbitrary small positive number $\epsilon$ we have

$$\sqrt{\frac{\alpha}{1-\alpha}}\le\sup_{X\in V(0,1)}\mathrm{VaR}_\alpha^+[X]\le\sup_{X\in V(0,1)}\mathrm{VaR}_{\alpha+\epsilon}[X]\le\sqrt{\frac{\alpha+\epsilon}{1-\alpha-\epsilon}},$$

from which we get

$$\sup_{X\in V(0,1)}\mathrm{VaR}_\alpha^+[X]=\sqrt{\frac{\alpha}{1-\alpha}}.$$

Note that for the arbitrary small $0<\epsilon<\alpha$,

$$\sup_{X\in V(0,1)}\mathrm{VaR}_{\alpha-\epsilon}^+[X]\le\sup_{X\in V(0,1)}\mathrm{VaR}_\alpha[X]\le\sup_{X\in V(0,1)}\mathrm{VaR}_\alpha^+[X].$$

Thus, we have

$$\sqrt{\frac{\alpha-\epsilon}{1-\alpha+\epsilon}}\le\sup_{X\in V(0,1)}\mathrm{VaR}_\alpha[X]\le\sqrt{\frac{\alpha}{1-\alpha}},$$

from which we get

$$\sup_{X\in V(0,1)}\mathrm{VaR}_\alpha[X]=\sqrt{\frac{\alpha}{1-\alpha}}.$$

Moreover, $\sup_{X\in V(0,1)}\mathrm{VaR}_\alpha[X]$ is not attainable, $\sup_{X\in V(0,1)}\mathrm{VaR}_\alpha^+[X]$ is attainable at the two-point distribution:

$$P\left(X_*=\sqrt{\frac{\alpha}{1-\alpha}}\right)=1-\alpha,\ P\left(X_*=-\sqrt{\frac{1-\alpha}{\alpha}}\right)=\alpha.$$

Statement (ii) can then be proved by using (i) and formula $F_Z^{-1}(\alpha)=-F_{-Z}^{-1+}(1-\alpha)$ , $\alpha\in(0,1)$. This ends the proof. □

**Remark 3.1** *The result for* $\sup_{X\in V(\mu,\sigma)} \mathrm{VaR}_\alpha[X]$ *was first studied in El Ghaoui et al. (2003) by using a conic programming approach, see also Li (2018). Li (2018) established the result for* $\sup_{X\in V(\mu,\sigma)} \mathrm{VaR}^+_\alpha[X]$ *based on the theory of stochastic comparison and joint mixability (Note that they referred to* $\mathrm{VaR}_\alpha[X]$ *as the right-continuous VaR). The value* $\sup_{X\in V(\mu,\sigma)} \mathrm{VaR}_\alpha[X]$ *is not attainable was noted by Cai et al. (2025), see also Bernard et al. (2024).*

### 3.2 Case of symmetric distributions

The family of symmetric distributions (about the expectation $\mu$) is characterized by equivalent relations

$$F(x-\mu)+F((\mu-x)-)=1,\ F^{-1}(p)+F^{-1+}(1-p)=2\mu,$$

and

$$P(X\le x)=P(X\ge 2\mu-x).$$

The following result is called Bienaymé-Chebyshev's inequality.

**Lemma 3.2** *(Ion et al. 2023) Let X be a symmetric random variable with mean* $\mu$ *and variance* $\sigma^2$*. Then, for any* $x\ge 0$*, with* $w=\max\{1,x\}$*, the inequality*

$$P\left(\frac{X-\mu}{\sigma}\ge x\right)\le\frac{1}{2w^2}$$

*holds, with equality if*

$$P\left(\frac{X-\mu}{\sigma}=-w\right)=P\left(\frac{X-\mu}{\sigma}=w\right)=\frac{1}{2w^2},\ P\left(X=\mu\right)=1-\frac{1}{w^2}.$$

Using Lemma 3.2 and the similar steps of analysis as those of Proposition 3.1, we obtain the following proposition.

**Proposition 3.2** *Let X be a symmetric random variable with mean* $\mu$ *and variance* $\sigma^2$ *. (i) We have that*

$$\sup_{X\in V_S(\mu,\sigma)} \mathrm{VaR}^+_\alpha[X]=\sup_{X\in V_S(\mu,\sigma)} \mathrm{VaR}_\alpha[X]=\begin{cases}\mu+\sigma\sqrt{\frac{1}{2(1-\alpha)}}, & \text{if } \alpha\ge\frac{1}{2},\\ \mu, & \text{if } \alpha<\frac{1}{2}.\end{cases} \tag{3.3}$$

*For* $\alpha\ge\frac{1}{2}$*, the first supremum in (3.3) is attained by the three-point distribution:* $P(X_*=\mu)=2\alpha-1$ *and*

$$P\left(X_* = \mu - \sigma\sqrt{\frac{1}{2(1-\alpha)}}\right) = P\left(X_* = \mu + \sigma\sqrt{\frac{1}{2(1-\alpha)}}\right) = 1 - \alpha.$$

*(ii) We have that*

$$\inf_{X \in V_S(\mu,\sigma)} \mathrm{VaR}_\alpha[X] = \inf_{X \in V_S(\mu,\sigma)} \mathrm{VaR}_\alpha^+[X] = \begin{cases} \mu - \sigma\sqrt{\frac{1}{2\alpha}}, & \text{if } 0 < \alpha \le \frac{1}{2}, \\ \mu, & \text{if } \alpha > \frac{1}{2}. \end{cases} \tag{3.4}$$

*For $0 < \alpha \le \frac{1}{2}$, the first infimum in (3.4) is attained by the three-point distribution: $P(X^* = \mu) = 1 - 2\alpha$ and*

$$P\left(X^* = \mu - \sigma\sqrt{\frac{1}{2\alpha}}\right) = P\left(X^* = \mu + \sigma\sqrt{\frac{1}{2\alpha}}\right) = \alpha.$$

**Remark 3.2** *(Li 2018) established the result for $\sup_{X \in V_S(\mu,\sigma)} \mathrm{VaR}_\alpha^+[X]$ based on the theory of stochastic comparison. Symmetry rules out the asymmetric two–point distribution that attains the bound in Proposition 3.1; the extremal distribution then becomes a three-point distribution with equal tail masses, which is the unique symmetric configuration with the given moments that maximizes a quantile.*

### 3.3 Case of unimodal distributions

We will use the following Vysochanskiĭ and Petunin's inequality.

**Lemma 3.3** *(Ion et al. 2023) Let the distribution of the standardized random variable Z be unimodal. Then, for any $v \ge 0$, the inequality*

$$P(Z \ge v) \le \begin{cases} \frac{3-v^2}{3(1+v^2)}, & \text{if } v \in \left[0, \sqrt{\frac{5}{3}}\right), \\ \frac{4}{9(1+v^2)}, & \text{if } v \in \left[\sqrt{\frac{5}{3}}, +\infty\right), \end{cases}$$

*holds, with equality for $v \in \left[0, \sqrt{\frac{5}{3}}\right)$ if $P(Z = v) = (3 - v^2)/(3(1 + v^2))$ and the rest of its mass, $4v^2/(3(1 + v^2))$, uniformly distributed on the interval $[-(3 + v^2)/2v, v]$, and with equality for $v \in \left[\sqrt{\frac{5}{3}}, +\infty\right)$ if*

*$P(Z = -1/v) = (3v^2 - 1)/(3(1 + v^2))$, and the rest of its mass, $4/(3(1 + v^2))$ , uniformly distributed on the interval $[-1/v, (1 + 3v^2)/(2v)]$.*

Using Lemma 3.3 and the similar steps of analysis as those of Proposition 3.1, we obtain the following proposition.

**Proposition 3.3** *Let X be a unimodal random variable with mean $\mu$ and variance $\sigma^2$ . Then, the following statements hold:*

*(i) We have, for $0 < \alpha < 1$,*

$$\sup_{X \in V_U(\mu,\sigma)} \mathrm{VaR}_\alpha^{\pm}[X] = \begin{cases} \mu + \sigma\sqrt{\frac{3\alpha}{4-3\alpha}}, & \text{if } \alpha \in \left[0, \frac{5}{6}\right), \\ \mu + \sigma\sqrt{\frac{4}{9(1-\alpha)} - 1}, & \text{if } \alpha \in \left[\frac{5}{6}, 1\right). \end{cases} \tag{3.5}$$

*For $\alpha \in \left[0, \frac{5}{6}\right)$, the value $\sup_{X\in V_U(\mu,\sigma)} \mathrm{VaR}_\alpha^+[X]$ is attained by the worst-case rv $X_*$ whose distribution is a mixture of the atom at $\mu + \sigma\sqrt{\frac{3\alpha}{4-3\alpha}}$ with probability $1 - \alpha$, and uniform distribution on interval*

$$\left[\mu - \sigma(2-\alpha)\sqrt{\frac{3}{\alpha(4-3\alpha)}}, \mu + \sigma\sqrt{\frac{3\alpha}{4-3\alpha}}\right]$$

*with probability $\alpha$; For $\alpha \in \left[\frac{5}{6}, 1\right)$, the value $\sup_{X\in V_U(\mu,\sigma)} \mathrm{VaR}_\alpha^+[X]$ is attained by the worst-case rv $X_*$ whose distribution is a mixture of the atom at $\mu - \sigma\sqrt{\frac{4}{9\alpha-5} - 1}$ with probability $3\alpha - 2$, and uniform distribution on interval*

$$\left[\mu - \sigma\sqrt{\frac{4}{9\alpha-5} - 1}, \mu + \sigma(3\alpha-1)\sqrt{\frac{1}{(1-\alpha)(9\alpha-5)}}\right]$$

*with probability $3(1-\alpha)$.*

*(ii) We have, for $0 < \alpha < 1$,*

$$\inf_{X \in V_U(\mu,\sigma)} \mathrm{VaR}_\alpha^{\pm}[X] = \begin{cases} \mu - \sigma\sqrt{\frac{3(1-\alpha)}{1+3\alpha}}, & \text{if } \alpha \in \left[\frac{1}{6}, 1\right), \\ \mu - \sigma\sqrt{\frac{4}{9\alpha} - 1}, & \text{if } \alpha \in \left[0, \frac{1}{6}\right). \end{cases} \tag{3.6}$$

*For $\alpha \in \left[\frac{1}{6}, 1\right)$, the value $\inf_{X\in V_U(\mu,\sigma)} \mathrm{VaR}_\alpha^-[X]$ is attained by the best-case rv $X^*$ whose distribution is a mixture of the atom at $\mu - \sigma\sqrt{\frac{3(1-\alpha)}{1+3\alpha}}$ with probability $\alpha$, and uniform distribution on interval*

$$\left[\mu - \sigma\sqrt{\frac{3(1-\alpha)}{1+3\alpha}}, \mu + \sigma(1+\alpha)\sqrt{\frac{3}{(1-\alpha)(1+3\alpha)}}\right]$$

*with probability $1-\alpha$; For $\alpha \in \left[0, \frac{1}{6}\right)$, the value $\inf_{X\in V_U(\mu,\sigma)} \mathrm{VaR}_\alpha^-[X]$ is attained by the best-case rv $X^*$ whose distribution is a mixture of the atom at $\mu + \sigma\sqrt{\frac{4}{4-9\alpha} - 1}$ with probability $1 - 3\alpha$, and uniform distribution on interval*

$$\left[\mu - \sigma(2-3\alpha)\sqrt{\frac{1}{\alpha(4-9\alpha)}}, \mu + \sigma\sqrt{\frac{4}{4-9\alpha} - 1}\right]$$

*with probability $3\alpha$.*

**Remark 3.3** *(Li 2018) established the result for* $\sup_{X\in V_U(\mu,\sigma)} \mathrm{VaR}_\alpha^+[X]$ *when* $\alpha \geq 5/6$*. Bernard et al. (2020) obtained* $\sup_{X\in\mathcal{A}_U(\mu,s)} \mathrm{VaR}_\alpha[X]$ *and* $\inf_{X\in\mathcal{A}_U(\mu,s)} \mathrm{VaR}_\alpha[X]$*, where*

$$\mathcal{A}_U(\mu, s) = \{X : X \,\text{is unimodal}, E[X] = \mu, \text{var}[X] \leq s^2\}.$$

*We remark that our results improve their corresponding results because*
$V_U(\mu,\sigma) \subseteq \mathcal{A}_U(\mu, s)$ *for* $\sigma \leq s$*. We also remark that* $\sup_{X\in V_U(\mu,\sigma)} \mathrm{VaR}_\alpha[X]$ *is not attainable.*

### 3.4 Case of symmetric and unimodal distributions

Taking both symmetry and unimodality constraints into account, we now establish the following result.

**Proposition 3.4** *Let $X$ be a symmetric and unimodal random variable with mean $\mu$ and variance $\sigma^2$. Then,*

*(i) we have that*

$$\sup_{X\in V_{US}(\mu,\sigma)} \mathrm{VaR}_\alpha^\pm[X] = \begin{cases} \mu, & \text{if } \alpha \in [0, \frac{1}{2}), \\ \mu + \sigma\sqrt{3}(2\alpha - 1), & \text{if } \alpha \in \left[\frac{1}{2}, \frac{5}{6}\right), \\ \mu + \sigma\sqrt{\frac{2}{9(1-\alpha)}}, & \text{if } \alpha \in [\frac{5}{6}, 1). \end{cases} \tag{3.7}$$

*For $\alpha \in \left[\frac{1}{2}, \frac{5}{6}\right)$, the value* $\sup_{X\in V_{US}(\mu,\sigma)} \mathrm{VaR}_\alpha^+[X]$ *is attained by the worst-case rv $X_*$ whose distribution is the uniform distribution on $[\mu - \sigma\sqrt{3}, \mu + \sigma\sqrt{3}]$; For $\alpha \in \left[\frac{5}{6}, 1\right)$, the value* $\sup_{X\in V_{US}(\mu,\sigma)} \mathrm{VaR}_\alpha^+[X]$ *is attained by the worst-case rv $X_*$ whose distribution is the mixture of the uniform distribution on*

$$\left[\mu - \sigma\sqrt{\frac{1}{2(1-\alpha)}}, \mu + \sigma\sqrt{\frac{1}{2(1-\alpha)}}\right]$$

*and the distribution degenerate at $\mu$ such that the point mass at $\mu$ equals $6\alpha - 5$.*

*(ii) We have that*

$$\inf_{X\in V_{US}(\mu,\sigma)} \mathrm{VaR}_\alpha^{\pm}[X] = \begin{cases} \mu, & \text{if } \alpha \in (\frac{1}{2}, 1), \\ \mu - \sigma\sqrt{3}(1-2\alpha), & \text{if } \alpha \in \left(\frac{1}{6}, \frac{1}{2}\right], \\ \mu - \sigma\sqrt{\frac{2}{9\alpha}}, & \text{if } \alpha \in (0, \frac{1}{6}]. \end{cases} \tag{3.8}$$

*For* $\alpha \in \left(\frac{1}{6}, \frac{1}{2}\right]$, *the value* $\inf_{X\in V_{US}(\mu,\sigma)} \mathrm{VaR}_\alpha^{-}[X]$ *is attained by the best-case rv* $X^*$ *whose distribution is the uniform distribution on* $[\mu - \sigma\sqrt{3}, \mu + \sigma\sqrt{3}]$*; For* $\alpha \in \left(0, \frac{1}{6}\right]$, *the value* $\inf_{X\in V_{US}(\mu,\sigma)} \mathrm{VaR}_\alpha^{-}[X]$ *is attained by the best-case rv* $X^*$ *whose distribution is the mixture of a uniform distribution on*

$$\left[\mu - \sigma\sqrt{\frac{1}{2\alpha}}, \mu + \sigma\sqrt{\frac{1}{2\alpha}}\right]$$

*and a distribution degenerate at* $\mu$ *such that the point mass at* $\mu$ *equals* $1 - 6\alpha$.

***Proof*** It is a direct consequence of the following lemma. □

**Lemma 3.4** *(Ion et al. 2023) If the distribution of the standardized random variable* $Z$ *is symmetric and unimodal, then*

$$P(Z \geq v) \leq \begin{cases} \frac{1}{2}(1 - \frac{v}{\sqrt{3}}), & \text{if } v \in \left[0, \frac{2}{\sqrt{3}}\right), \\ \frac{4}{9}\frac{1}{2v^2}, & \text{if } v \in \left[\frac{2}{\sqrt{3}}, +\infty\right), \end{cases}$$

*holds. Equality is attained by the mixture of a uniform distribution on* $[-\sqrt{3} \vee (\frac{3}{2}v), \sqrt{3} \vee (\frac{3}{2}v)]$ *and a distribution degenerate at 0 such that the point mass at 0 equals* $[1 - \frac{4}{3v^2}] \vee 0$.

**Remark 3.4** *(Li 2018) established* $\sup_{X\in V_{US}(\mu,\sigma)} \mathrm{VaR}_\alpha^{+}[X]$ *when* $\alpha \geq 5/6$ *. (Bernard et al. 2020, 2025a) obtained* $\sup_{X\in \mathcal{A}_{US}(\mu,s)} \mathrm{VaR}_\alpha[X]$, *where* $\mathcal{A}_{US}(\mu, s) = \{X : X \text{ is unimodal and symmetric}, E[X] = \mu, \mathrm{var}[X] \leq s^2\}$. *We remark that our results improved the corresponding results in (Bernard et al. 2022, 2025a) because* $V_{US}(\mu,\sigma) \subseteq \mathcal{A}_{US}(\mu, s)$ *for* $\sigma \leq s$.

**Remark 3.5** *The tightening from the general case to the symmetric-unimodal case is driven by the progressive elimination of admissible tail configurations: without shape constraints, mass can concentrate unilaterally in a single tail; symmetry forces bilateral balance; unimodality restricts the distribution to at most one peak and forces interior mass to be continuous; their combination further requires the distribution to be symmetric about the mode.*

## 4 Bounds of distortion risk measures

The following two lemmas are useful in proving our main result.

**Lemma 4.1** *(Dhaene et al. 2012) For any distortion function h and* $X \in L^{\infty}$ *, the distortion risk measure* $\rho_h[X]$ *has the following Lebesgue-Stieltjes integral representation:*

*(i) If h is right-continuous, then*

$$\rho_h[X] = \int_0^1 F_X^{-1+}(1-p)dh(p) = \int_0^1 F_X^{-1+}(p)d\tilde{h}(p);$$

*(ii) If h is left-continuous, then*

$$\rho_h[X] = \int_0^1 F_X^{-1}(1-p)dh(p) = \int_0^1 F_X^{-1}(p)d\tilde{h}(p);$$

*(iii) If h is continuous, then*

$$\rho_h[X] = \int_0^1 F_X^{-1}(1-p)dh(p) = \int_0^1 F_X^{-1}(p)d\tilde{h}(p),$$

*where* $\tilde{h}(p) = 1 - h(1-p)$ $(p \in [0, 1])$ *is a dual distortion function.*

**Lemma 4.2** *(Modified Schwarz inequality, Moriguti (1953)) Let H be a function of bounded variation in the closed interval [a, b] and continuous at both ends. Then, the inequality*

$$\int_a^b x(t)dH(t) \leq \int_a^b x(t)\bar{h}(t)dt \tag{4.1}$$

*holds for any nondecreasing function x(t) for which the integrals exist and are finite, where* $\bar{h}(t)$ *is the right-hand derivative of the greatest convex minorant* $\bar{H}$ *of H. The equality in (4.1) holds if and only if x(t) is a constant in every interval contained in* $\{t : \bar{H}(t) < \min\{H(t-0), H(t+0)\}\}$ *and, at every point of discontinuity, if any, of H(t),* $x(t_n) = x(t_n + 0)$ *when* $H(t_n - 0) < H(t_n + 0)$, $x(t_n) = x(t_n - 0)$ *when* $H(t_n - 0) > H(t_n + 0)$.

## 4.1 General distributions

We now deal with the extreme-case distortion risk measures (DRMs) for the univariate case under the assumption that the first two moments are known. Li (2018) and Liu et al. (2020) obtained $\sup_{X \in V(\mu,\sigma)} \rho_h[X]$ and the worst-case distribution when $h$ is a concave distortion function. Shao and Zhang (2023) derived the worst-case DRMs under left-continuous distortion function, and derived the best-case DRMs under right-continuous distortion function. In addition, they also found the corresponding distributions. ( Bernard et al. (2024), Corollary 3.9) derived the worst-case DRMs in which the increasing and absolutely continuous distortion functions were considered. Cai et al. (2025) obtained $\sup_{X \in V(\mu,\sigma)} \rho_h[X]$ for the general $h$, when $h$ is left-continuous, the form of worst distribution was found without determining coefficients. Note that the definition of $\rho_h[X]$ in Shao and Zhang (2023) and Cai et al. (2025) is equivalent to $\rho_{\tilde{h}}[X]$ with $\tilde{h}(t) = 1 - h(1-t)$ in this paper.

**Proposition 4.1** *Let $\mu \in \mathbb{R}$ and $\sigma > 0$. Then, the following statements hold:*

*(i) For any distortion function h, we have*

$$\sup_{X \in V(\mu,\sigma)} \rho_h[X] = \mu + \sigma \sqrt{\int_0^1 (\tilde{h}'_*(p) - 1)^2 dp}, \tag{4.2}$$

*where $\tilde{h}'_*$ denotes the right derivative of $\tilde{h}_*$. Moreover, if h is right-continuous, then, above supremum is attainable at the worst-case random variable $X_* = \arg\sup_{X \in V(\mu,\sigma)} \rho_h[X]$ with*

$$F_{X_*}^{-1+}(p) = \begin{cases} \mu, & \text{if } \tilde{h}'_*(p) = 1 \ (a.e.), \\ \mu + \sigma \frac{\tilde{h}'_*(p) - 1}{\sqrt{\int_0^1 (\tilde{h}'_*(p)-1)^2 dp}}, & \text{if } \tilde{h}'_*(p) \neq 1 \ (a.e.). \end{cases}$$

*(ii) For any distortion function h, we have*

$$\inf_{X \in V(\mu,\sigma)} \rho_h[X] = \mu - \sigma \sqrt{\int_0^1 (h'_*(p) - 1)^2 dp}, \tag{4.3}$$

*where $h'_*$ denotes the left derivative of $h_*$. Moreover, if h is left-continuous, then, the above infimum is attainable at the best-case random variable $X^* = \arg\inf_{X \in V(\mu,\sigma)} \rho_h[X]$ with*

$$F_{X^*}^{-1+}(p) = \begin{cases} \mu, & \text{if } h'_*(p) = 1 \ (a.e.), \\ \mu - \sigma \frac{h'_*(p) - 1}{\sqrt{\int_0^1 (h'_*(p)-1)^2 dp}}, & \text{if } h'_*(p) \neq 1 \ (a.e.). \end{cases}$$

***Proof*** We only give the proof of the statement (i) for the right-continuous $h$, the proof for the left-continuous $h$ is similar. If $\tilde{h}'_*(p) = 1$ (a.e.), the proof is trivial. Next, we assume $\tilde{h}'_*(p) \neq 1$ (a.e.). Using Lemmas 4.1 and 4.2, for any constant $c$, we have

$$\begin{aligned} \rho_h[X] =& \mu + \int_0^1 (F_X^{-1+}(p) - \mu) d\tilde{h}(p) \leq \mu + \int_0^1 (F_X^{-1+}(p) - \mu) \tilde{h}'_*(p) dp \\ =& \mu + \int_0^1 (F_X^{-1+}(p) - \mu)(\tilde{h}'_*(p) - c) dp \\ \leq& \mu + \left( \int_0^1 (F_X^{-1+}(p) - \mu)^2 dp \right)^{\frac{1}{2}} \cdot \left( \int_0^1 (\tilde{h}'_*(p) - c)^2 dp \right)^{\frac{1}{2}} \\ =& \mu + \sigma \left( \int_0^1 (\tilde{h}'_*(p) - c)^2 dp \right)^{\frac{1}{2}}, \end{aligned}$$

with the last inequality become equality if

$$F_{X_*}^{-1+}(p) = \mu + \sigma \frac{\tilde{h}'_*(p) - 1}{\sqrt{\int_0^1 (\tilde{h}'_*(p) - 1)^2 dp}}.$$

Note that

$$\inf_{c \in \mathbb{R}} \left( \int_0^1 (\tilde{h}'_*(p) - c)^2 dp \right)^{\frac{1}{2}} = \sqrt{\int_0^1 (\tilde{h}'_*(p) - 1)^2 dp},$$

and

$$\int_0^1 F_{X_*}^{-1+}(p) d\tilde{h}(p) = \int_0^1 F_{X_*}^{-1+}(p) \tilde{h}'_*(p) dp.$$

Therefore,

$$\rho_h[X] \leq \mu + \sigma \sqrt{\int_0^1 (\tilde{h}'_*(p) - 1)^2 dp}$$

with equality holding if

$$F_{X_*}^{-1+}(p) = \mu + \sigma \frac{\tilde{h}'_*(p) - 1}{\sqrt{\int_0^1 (\tilde{h}'_*(p) - 1)^2 dp}}.$$

Hence, (i) follows. The statement (ii) can be verified directly from (i) and the relationship

$$\inf_{X \in V(\mu,\sigma)} \rho_h[X] = \mu - \sigma \sup_{Z \in V(0,1)} \int_0^1 F_{-Z}^{-1+}(p) dh(p).$$

This completes the proof. □

**Remark 4.1** *The argument adopted in Shao and Zhang (2023) and Cai et al. (2025) requiring nontrivial technical proofs.*

*The method used here is different from the one in the literature and the proof presented here is much shorter.*

### 4.2 Symmetric distributions

Taking symmetry constraint into account, we have the following result. In the approach, we are greatly inspired by the recent work of Shao and Zhang (2023).

**Proposition 4.2** *Let $\mu \in \mathbb{R}$ and $\sigma > 0$. Then, the following statements hold:*

*(i) For any distortion function h, we have*

$$\sup_{X\in V_S(\mu,\sigma)} \rho_h[X] = \mu + \frac{1}{2}\sigma\sqrt{\int_0^1 (\tilde{h}'_*(p) - \tilde{h}'_*(1-p))^2 dp}, \tag{4.4}$$

*where $\tilde{h}'_*$ denotes the right derivative of $\tilde{h}_*$. Moreover, if h is right-continuous, then, above supremum is attainable at the worst-case random variable $X_* = \arg\sup_{X\in V_S(\mu,\sigma)} \rho_h[X]$ with*

$$F_{X_*}^{-1+}(p) = \begin{cases} \mu, & \text{if } \tilde{h}'_*(p) - \tilde{h}'_*(1-p) = 0 \ (a.e.), \\ \mu + \sigma\frac{\tilde{h}'_*(p)-\tilde{h}'_*(1-p)}{\sqrt{\int_0^1(\tilde{h}'_*(p)-\tilde{h}'_*(1-p))^2dp}}, & \text{if } \tilde{h}'_*(p) - \tilde{h}'_*(1-p) \neq 0 \ (a.e.). \end{cases}$$

*(ii) For any distortion function h, we have*

$$\inf_{X\in V_S(\mu,\sigma)} \rho_h[X] = \mu - \frac{1}{2}\sigma\sqrt{\int_0^1 (h'_*(p) - h'_*(1-p))^2 dp}. \tag{4.5}$$

*Moreover, if h is left-continuous, then, the above infimum is attainable at the best-case random variable $X^* = \arg\inf_{X\in V_S(\mu,\sigma)} \rho_h[X]$ with*

$$F_{X^*}^{-1+}(p) = \begin{cases} \mu, & \text{if } h'_*(p) - h'_*(1-p) = 0 \ (a.e.), \\ \mu - \sigma\frac{h'_*(p)-h'_*(1-p)}{\sqrt{\int_0^1(h'_*(p)-h'_*(1-p))^2dp}}, & \text{if } h'_*(p) - h'_*(1-p) \neq 0 \ (a.e.). \end{cases}$$

***Proof*** Using $F_X^{-1+}(p) + F_X^{-1+}(1-p) = 2\mu$ ($p \in (0,1)$, *a.e.*) for $X \in V_S(\mu,\sigma)$, we express $\rho_h[X]$ as

$$\rho_h[X] = \mu + \frac{1}{2}\int_0^1 \left(F_X^{-1+}(p) - \mu\right) d(\tilde{h}(p) + \tilde{h}(1-p)),$$

from which we get the desired results by using same arguments as those for Proposition 4.1. □

**Remark 4.2** *Shao and Zhang (2023) established the results for dual DRMs. To prove the main theorem, they undertook a logical progression consisting of three long distinct steps. But, the proof provided here is short and simple relying mainly on Lemma 4.2.*

### 4.3 Unimodal distributions

Li (2018) discussed RVaR $\sup_{X\in V_U(\mu,\sigma)} \mathrm{RVaR}_{\alpha,\beta}[X]$ with $\frac{5}{6}\le\alpha\le\beta<1$ for unimodal distributions. Bernard et al. (2020) obtained the results for $\sup_{X\in\mathcal{A}_U(\mu,s)} \mathrm{RVaR}_{\alpha,\beta}[X]$ and $\inf_{X\in\mathcal{A}_U(\mu,s)} \mathrm{RVaR}_{\alpha,\beta}[X]$, where $\mathcal{A}_U(\mu,s)=\{X : X \text{ is unimodal}, E[X]=\mu, \mathrm{var}[X]\le s^2\}$. In particular, the $\mathrm{VaR}_\alpha$ and $\mathrm{TVaR}_\alpha$ bounds can be derived from the results of $\mathrm{RVaR}_{\alpha,\beta}[X]$. Inspired by Li (2018) and Bernard et al. (2020), we proceed to study the extreme-case DRMs with general distortion functions under the unimodality constraint. From Theorem 2 in Bernard et al. (2020), the result for $\sup_{X\in\mathcal{A}_U(\mu,s)} \mathrm{RVaR}_\alpha[X]$, we can obtain an improved version for $\sup_{X\in\mathcal{A}_U(\mu,s)} \mathrm{TVaR}_\alpha[X]$ (see Corollary 4 in Bernard et al. (2020)). This result, which is summarized in the following lemma, plays an important role in the proof of Theorem 4.1.

**Lemma 4.3** *Let $\mu\in\mathbb{R}$ and $\sigma>0$. Then,*

$$\sup_{X\in V_U(\mu,\sigma)} \mathrm{TVaR}_\alpha[X]=\begin{cases}\mu+\sigma\frac{\sqrt{\alpha(8/9-\alpha)}}{1-\alpha}, & \text{if } 0\le\alpha<\frac{1}{2},\\ \mu+\sigma\sqrt{\frac{8}{9(1-\alpha)}-1}, & \text{if } \frac{1}{2}\le\alpha<1.\end{cases} \tag{4.6}$$

*Moreover, for $\alpha\in\left[\frac{1}{2},1\right)$, the supremum is attained by the worst-case rv $X_*$ with*

$$F^{-1}_{X_*}(p)=\begin{cases}\mu-3\sigma\sqrt{\frac{1-\alpha}{9\alpha-1}}, \text{ if } p\in[0,\frac{3\alpha-1}{2}),\\ \mu+\frac{8\sigma\left(p-\frac{3\alpha-1}{2}\right)}{3\sqrt{(1-\alpha)^3(9\alpha-1)}}-3\sigma\sqrt{\frac{1-\alpha}{9\alpha-1}}, \text{ if } p\in[\frac{3\alpha-1}{2},1);\end{cases}$$

*For $\alpha\in\left[0,\frac{1}{2}\right)$, the supremum is attained by the worst-case rv $X_*$ with*

$$F^{-1}_{X_*}(p)=\begin{cases}\mu+\frac{8\sigma}{3}\frac{p-\frac{3\alpha}{2}}{\sqrt{\alpha(8-9\alpha)}}+3\sigma\sqrt{\frac{\alpha}{8-9\alpha}}, \text{ if } p\in[0,\frac{3\alpha}{2}),\\ \mu+3\sigma\sqrt{\frac{\alpha}{8-9\alpha}}, \text{ if } p\in[\frac{3\alpha}{2},1).\end{cases}$$

We shall use the following two notations:

$$\begin{aligned}\Delta_R(g,b)=&\frac{-(1+b^2)}{\sqrt{(1-b)^3(1/3+b)}}+\frac{2b}{\sqrt{(1-b)^3(1/3+b)}}g(b)\\ &+\frac{2}{\sqrt{(1-b)^3(1/3+b)}}\int_b^1 pdg(p)\end{aligned}$$

and

$$\Delta_L(g,b) = \sqrt{\frac{3b}{4-3b}} - \frac{2b\sqrt{3}}{\sqrt{b^3(4-3b)}} g(b) + \frac{2\sqrt{3}}{\sqrt{b^3(4-3b)}} \int_0^b p dg(p).$$

We first consider the worst case.

**Theorem 4.1** *Assume that h is a distortion function and denote by* $\tilde{h}(p) = 1 - h(1-p)$.

*(i) If h is a simple function, then,*

$$\sup_{X \in V_U(\mu,\sigma)} \rho_h[X] = \mu + \sigma \int_0^{\frac{5}{6}} \sqrt{\frac{3p}{4-3p}} d\tilde{h}(p) + \sigma \int_{\frac{5}{6}}^1 \sqrt{\frac{4}{9(1-p)} - 1} d\tilde{h}(p). \tag{4.7}$$

*(ii) If h is a concave function, then,*

$$\sup_{X \in V_U(\mu,\sigma)} \rho_h[X] = \mu + \frac{1}{3}\sigma \int_0^{\frac{1}{2}} \sqrt{p(8-9p)} d\tilde{h}'(p) + \frac{1}{3}\sigma \int_{\frac{1}{2}}^1 \sqrt{(1-p)(9p-1)} d\tilde{h}'(p), \tag{4.8}$$

*where* $\tilde{h}'(p)$ *is the right derivative of* $\tilde{h}(p)$.

*(iii) If h is a distortion function with* $h(0+) = 0$ *and* $h(1-) = 1$*, then*

$$\mu + \sigma \sup_{b \in [0,1]} \left\{ \Delta_R(\tilde{h}, b), \Delta_L(\tilde{h}, b) \right\} \le \sup_{X \in V_U(\mu,\sigma)} \rho_h[X] \le \mu + \frac{1}{3}\sigma \int_0^{\frac{1}{2}} \sqrt{p(8-9p)} d(\widetilde{h^*})'(p) + \frac{1}{3}\sigma \int_{\frac{1}{2}}^1 \sqrt{(1-p)(9p-1)} d(\widetilde{h^*})'(p), \tag{4.9}$$

*where* $h^*$ *is the concave envelope of h.*

***Proof*** (i) We only give the proof of the statement (i) for the right-continuous $h$, the proof for the case of left-continuous $h$ is similar. Without loss of generality, we assume $\tilde{h}$ has the form

$$\tilde{h}(t) = c_1 1_{\{t_0\}}(t) + \sum_{i=1}^{n} c_i 1_{(t_{i-1},t_i]}(t),$$

where $0 = c_1 \leq c_2 \leq \cdots \leq c_n = 1$ are constants, $0 = t_0 \leq t_1 \leq \cdots \leq t_n = 1$. Then,

$$\rho_h[X] = \sum_{i=1}^{n-1} (c_{i+1} - c_i) F_X^{-1+}(t_i).$$

Since *n* is a finite integer, all summands have the same monotonicity, so the interchangeability of the supremum and summation is allowed, we have

$$\begin{aligned}\sup_{X \in V_U(\mu,\sigma)} \rho_h[X] &= \sup_{X \in V_U(\mu,\sigma)} \sum_{i=1}^{n-1} (c_{i+1} - c_i) F_X^{-1+}(t_i) \\ &= \sum_{i=1}^{n-1} (c_{i+1} - c_i) \sup_{X \in V_U(\mu,\sigma)} F_X^{-1+}(t_i),\end{aligned}$$

this, together with (3.5), yields (4.7).

(ii) We first assume *h* is a piecewise linear concave function. Then,

$$\begin{aligned}\sup_{X \in V_U(\mu,\sigma)} \rho_h[X] =& \mu + \sigma \sup_{Z \in V_U(0,1)} \int_0^1 F_Z^{-1+}(p) d\tilde{h}(p) \\ =& \mu + \sigma \sup_{Z \in V_U(0,1)} \int_0^1 (1-p) \mathrm{TVaR}_p[Z] d\tilde{h}'(p),\end{aligned}$$

where $\tilde{h}'(p) \geq 0$ is the right derivative of the piecewise linear convex $\tilde{h}(p)$. Obviously, $\tilde{h}'(p)$ is a simple function. It follows that

$$\sup_{X \in V_U(\mu,\sigma)} \rho_h[X] = \mu + \sigma \int_0^1 (1-p) \sup_{Z \in V_U(0,1)} \mathrm{TVaR}_p[Z] d\tilde{h}'(p).$$

The result (4.8) follows easily by using the result (4.6).

Next, we consider any concave distortion function *h*. Note that there exists a sequence of concave piecewise linear distortion functions $h_1(x) \leq h_2(x) \leq \cdots \leq h_n(x) \leq \cdots \leq h(x)$ such that $h(x) = \lim_{n\to\infty} h_n(x)$. From the monotone convergence theorem we have $\lim_{n\to\infty} \rho_{h_n}[X] = \rho_h[X]$. It follows from the monotonicity of $\rho_{h_n}[X]$, we have

$$\begin{aligned}\sup_{X\in V_U(\mu,\sigma)}\rho_h[X]=&\sup_{X\in V_U(\mu,\sigma)}\lim_{n\to\infty}\rho_{h_n}[X]=\lim_{n\to\infty}\sup_{X\in V_U(\mu,\sigma)}\rho_{h_n}[X]\\ &=\lim_{n\to\infty}\left(\mu+\sigma\int_0^1(1-p)\sup_{Z\in V_U(0,1)}\mathrm{TVaR}_p[Z]d\tilde{h}_n'(p)\right)\\ &=\mu+\sigma\int_0^1(1-p)\sup_{Z\in V_U(0,1)}\mathrm{TVaR}_p[Z]d\tilde{h}'(p),\end{aligned}$$

as desired.

(iii) The idea of Bernard et al. (2020) is particularly useful in proving our result. For $0\le b\le 1$, we introduce two sets $U_R$ and $U_L$ as follows:

$$U_R=\{X: F_X^{-1}(p)=a \text{ for } p\in[0,b); c(p-b)+a \text{ for } p\in[b,1]\}$$

and

$$U_L=\{X: F_X^{-1}(p)=a \text{ for } p\in[b,1]; c(p-b)+a \text{ for } p\in[0,b)\}.$$

It is clear that

$$U_R\cap V(\mu,\sigma)\subseteq V_U(\mu,\sigma),\ \ U_L\cap V(\mu,\sigma)\subseteq V_U(\mu,\sigma).$$

Moreover, by moments matching, we have

$$U_R\cap V(\mu,\sigma)=\left\{X: F_X^{-1}(p)=\left\{\begin{array}{l}\mu-\sigma\sqrt{\frac{1-b}{1/3+b}},\ p\in[0,b),\\ \mu+\sigma\frac{2p-1-b^2}{\sqrt{(1-b)^3(1/3+b)}},\ p\in[b,1),\end{array}\right.\right\}$$

and

$$U_L\cap V(\mu,\sigma)=\left\{X: F_X^{-1}(p)=\left\{\begin{array}{l}\mu+\sigma\sqrt{3}\frac{2p-2b+b^2}{\sqrt{b^3(4-3b)}},\ p\in[0,b),\\ \mu+\sigma\sqrt{\frac{3b}{4-3b}},\ p\in[b,1),\end{array}\right.\right\}.$$

For $X\in U_R\cap V(\mu,\sigma)$, since the left-continuous and right-continuous inverses of a unimodal df coincide on (0, 1) (see, e.g., Bernard et al. (2025a)), we have

$$
\begin{aligned}
\rho_h[X] =&\mu + \int_0^1 (F_X^{-1}(p) - \mu) d\tilde{h}(p) \\
=&\mu - \sigma \int_0^b \sqrt{\frac{1-b}{1/3+b}} d\tilde{h}(p) + \sigma \int_b^1 \frac{2p - 1 - b^2}{\sqrt{(1-b)^3(1/3+b)}} d\tilde{h}(p) \\
=&\mu - \frac{\sigma(1+b^2)}{\sqrt{(1-b)^3(1/3+b)}} + \frac{2b\sigma}{\sqrt{(1-b)^3(1/3+b)}} \tilde{h}(b) \\
&+ \frac{2\sigma}{\sqrt{(1-b)^3(1/3+b)}} \int_b^1 p d\tilde{h}(p) \\
\equiv&\mu + \sigma \Delta_R(\tilde{h}, b).
\end{aligned}
\tag{4.10}
$$

Similarly, for $X \in U_L \cap V(\mu, \sigma)$, we have

$$
\begin{aligned}
\rho_h[X] =&\mu + \int_0^1 (F_X^{-1}(p) - \mu) d\tilde{h}(p) \\
=&\mu + \sigma \int_b^1 \sqrt{\frac{3b}{4-3b}} d\tilde{h}(p) + \sigma\sqrt{3} \int_0^b \frac{2p - 2b + b^2}{\sqrt{b^3(4-3b)}} d\tilde{h}(p) \\
=&\mu + \sigma \sqrt{\frac{3b}{4-3b}} - \frac{2b\sqrt{3}\sigma}{\sqrt{b^3(4-3b)}} \tilde{h}(b) \\
&+ \frac{2\sqrt{3}\sigma}{\sqrt{b^3(4-3b)}} \int_0^b p d\tilde{h}(p) \\
\equiv&\mu + \sigma \Delta_L(\tilde{h}, b).
\end{aligned}
\tag{4.11}
$$

Combining (4.10) with (4.11), we get

$$
\sup_{X \in V_U(\mu,\sigma)} \rho_h[X] \geq \mu + \sigma \sup_{b \in [0,1]} \left\{ \Delta_R(\tilde{h}, b), \Delta_L(\tilde{h}, b) \right\},
$$

which establishes the lower bound of (4.9). To obtain the upper bound of (4.9), it is well known that for any concave distortion function $h^*$, there exists a sequence of concave piecewise linear distortion functions $h_1^*(x) \leq h_2^*(x) \leq \cdots \leq h_n^*(x) \leq \cdots \leq h^*(x)$ such that $h^*(x) = \lim_{n\to\infty} h_n^*(x)$. From the monotone convergence theorem we have $\lim_{n\to\infty} \rho_{h_n^*}[X] = \rho_{h^*}[X]$. It follows from the monotonicity of $\rho_{h_n^*}[X]$ and note that $h \leq h^*$, we have

$$\begin{aligned}\sup_{X\in V_U(\mu,\sigma)} \rho_h[X] \le \sup_{X\in V_U(\mu,\sigma)} \rho_{h^*}[X] &= \sup_{X\in V_U(\mu,\sigma)} \lim_{n\to\infty} \rho_{h_n^*}[X] \\ &= \lim_{n\to\infty} \sup_{X\in V_U(\mu,\sigma)} \rho_{h_n^*}[X] \\ &= \lim_{n\to\infty}\left(\mu+\sigma\int_0^1 (1-p) \sup_{Z\in V_U(0,1)} \mathrm{TVaR}_p[Z] d(\widetilde{h_n^*})'(p)\right) \\ &= \mu+\sigma\int_0^1 (1-p) \sup_{Z\in V_U(0,1)} \mathrm{TVaR}_p[Z] d(\widetilde{h^*})'(p),\end{aligned}$$

from which and (4.6) we get the desired result. This completes the proof of Theorem 4.1. □

The next theorem considers the best case.

**Theorem 4.2** *Assume that h is a distortion function. Then, the following statements hold:*

*(i) If h is a simple function, then,*

$$\inf_{X\in V_U(\mu,\sigma)} \rho_h[X] = \mu - \sigma\left(\int_0^{\frac{5}{6}} \sqrt{\frac{3p}{4-3p}} dh(p) + \int_{\frac{5}{6}}^1 \sqrt{\frac{4}{9(1-p)}-1} dh(p)\right).$$

*(ii) If h is a convex function, then,*

$$\inf_{X\in V_U(\mu,\sigma)} \rho_h[X] = \mu - \frac{1}{3}\sigma\left(\int_0^{\frac{1}{2}} \sqrt{p(8-9p)} dh'(p) + \int_{\frac{1}{2}}^1 \sqrt{(1-p)(9p-1)} dh'(p)\right).$$

*(iii) If h is a distortion function with $h(0+)=0$ and $h(1-)=1$, then*

$$\begin{aligned}\mu - \frac{1}{3}\sigma\left(\int_0^{\frac{1}{2}} \sqrt{p(8-9p)} d\widetilde{(\tilde{h})^*}'(p) + \int_{\frac{1}{2}}^1 \sqrt{(1-p)(9p-1)} d\widetilde{(\tilde{h})^*}'(p)\right) \\ \le \inf_{X\in V_U(\mu,\sigma)} \rho_h[X] \\ \le \mu - \sigma \sup_{b\in[0,1]} \{\Delta_R(h,b), \Delta_L(h,b)\}.\end{aligned}$$

***Proof*** Using the formula

$$
\begin{aligned}
\inf_{X\in V_U(\mu,\sigma)} \rho_h[X] =& \mu + \sigma \inf_{Z\in V_U(0,1)} \int_0^1 F_Z^{-1}(p)d\tilde{h}(p) \\
=& \mu - \sigma \sup_{Z\in V_U(0,1)} \int_0^1 (-F_Z^{-1}(p))d\tilde{h}(p) \\
=& \mu - \sigma \sup_{Z\in V_U(0,1)} \int_0^1 F_{-Z}^{-1}(1-p)d\tilde{h}(p) \\
=& \mu - \sigma \sup_{Z\in V_U(0,1)} \int_0^1 F_Z^{-1}(p)dh(p) \\
=& \mu - \sigma \sup_{Z\in V_U(0,1)} \rho_{\tilde{h}}[Z],
\end{aligned}
$$

the result can be derived directly from Theorem 4.1. This ends the proof. □

### 4.4 Symmetric and unimodal distributions

Li (2018) discussed RVaR $\sup_{X\in V_{US}(\mu,\sigma)} \mathrm{RVaR}_{\alpha,\beta}[X]$ with $\frac{5}{6} \le \alpha \le \beta < 1$ for symmetric and unimodal distributions. Bernard et al. (2022, 2025a)) obtained $\sup_{X\in \mathcal{A}_{US}(\mu,s)} \mathrm{RVaR}_{\alpha,\beta}[X]$ and $\inf_{X\in \mathcal{A}_{US}(\mu,s)} \mathrm{RVaR}_{\alpha,\beta}[X]$, where $\mathcal{A}_{US}(\mu, s) = \{X : X \text{ is symmetric and unimodal}, E[X] = \mu, \mathrm{var}[X] \le s^2\}$. Inspired by Li (2018) and Bernard et al. (2020, 2025a), we investigate the extreme-case DRMs for symmetric and unimodal distributions with general distortion functions.

From Theorem 4.1 in Bernard et al. (2025a), we obtain an improved version for $\sup_{X\in \mathcal{A}_{US}(\mu,s)} \mathrm{TVaR}_\alpha[X]$ which plays an important role in the proof of Theorem 4.3.

**Lemma 4.4** *Let $\mu \in \mathbb{R}$ and $\sigma > 0$. Then,*

$$
\sup_{X\in V_{US}(\mu,\sigma)} \mathrm{TVaR}_\alpha[X] = \begin{cases} \mu + \sigma\frac{2\sqrt{\alpha}}{3(1-\alpha)}, & \text{if } \alpha \in (0, \frac{1}{3}), \\ \mu + \sigma\sqrt{3}\alpha, & \text{if } \alpha \in [\frac{1}{3}, \frac{2}{3}), \\ \mu + \frac{2}{3}\sigma\sqrt{\frac{1}{1-\alpha}}, & \text{if } \alpha \in [\frac{2}{3}, 1). \end{cases} \tag{4.12}
$$

*The supremum is attained by the worst-case rv $X_*$ with*

$$
F_{X_*}^{-1}(p) = \begin{cases} a + c(p-1+b), \ p \in (0, 1-b), \\ a, \ p \in [1-b, b], \\ a + c(p-b), \ p \in (b, 1), \end{cases}
$$

*in which a, b, and c are, respectively,*

*(I) $a = \mu$, $b = \frac{3}{2}\alpha - \frac{1}{2}$ and $c = \frac{2\sigma}{3(1-\alpha)^{3/2}}$ when $\alpha \in [\frac{2}{3}, 1)$;*
*(II) $a = \mu$, $b = \frac{1}{2}$ and $c = 2\sqrt{3}\sigma$ when $\alpha \in [\frac{1}{3}, \frac{2}{3})$;*
*(III) $a = \mu$, $b = 1 - \frac{3}{2}\alpha$ and $c = \frac{2\sigma}{3\alpha^{3/2}}$ when $\alpha \in (0, \frac{1}{3})$.*

Denote by $\Theta(g, b)$ and $\Upsilon(g)$ the following expressions

$$\Theta(g,b)=\frac{\int_0^{1-b}pdg(p)-(1-b)g(1-b)+\int_b^1 pdg(p)-b(1-g(b))}{\sqrt{\frac{2}{3}(1-b)^3}},$$

and

$$\Upsilon(g)=\frac{2}{3}\int_0^{\frac{1}{3}}\sqrt{p}dg'(p)+\sqrt{3}\int_{\frac{1}{3}}^{\frac{2}{3}}p(1-p)dg'(p)\\+\frac{2}{3}\int_{\frac{2}{3}}^{1}\sqrt{1-p}dg'(p).$$

**Theorem 4.3** *Let h be a distortion function and $\tilde{h}(p)=1-h(1-p)$. Then, the following statements hold.*

*(i) If h is a simple function, then,*

$$\sup_{X\in V_{US}(\mu,\sigma)}\rho_h[X]=\mu+\sqrt{3}\sigma\int_{\frac{1}{2}}^{\frac{5}{6}}(2p-1)d\tilde{h}(p)\\+\frac{\sqrt{2}}{3}\sigma\int_{\frac{5}{6}}^{1}\frac{1}{\sqrt{1-p}}d\tilde{h}(p). \tag{4.13}$$

*(ii) If h is a concave function, then,*

$$\sup_{X\in V_{US}(\mu,\sigma)}\rho_h[X]=\mu+\frac{2}{3}\sigma\int_0^{\frac{1}{3}}\sqrt{p}d\tilde{h}'(p)+\sqrt{3}\sigma\int_{\frac{1}{3}}^{\frac{2}{3}}p(1-p)d\tilde{h}'(p)\\+\frac{2}{3}\sigma\int_{\frac{2}{3}}^{1}\frac{1}{\sqrt{1-p}}d\tilde{h}'(p), \tag{4.14}$$

*where $\tilde{h}'(p)$ is the right derivative of $\tilde{h}(p)$.*

*(iii) If h is a general distortion function with $h(0+)=0$ and $h(1-)=1$, then*

$$\mu+\sigma\sup_{b\in[\frac{1}{2},1]}\Theta(\tilde{h},b)\le\sup_{X\in V_{US}(\mu,\sigma)}\rho_h[X]\le\mu+\sigma\Upsilon(\widetilde{h^*}), \tag{4.15}$$

*where $h^*$ is the concave envelope of h.*

***Proof*** See Appendix A1. □

The next theorem considers the best case.

**Theorem 4.4** *Let X be a symmetric unimodal random variable with mean $\mu$ and variance $\sigma^2$ and h be a left-continuous distortion function. Then, the following statements hold.*

*(i) If h is a simple function, then,*

$$\begin{aligned}\inf_{X\in V_{US}(\mu,\sigma)}\rho_h[X] =&\mu-\sqrt{3}\sigma\int_{\frac{1}{2}}^{\frac{5}{6}}(2p-1)dh(p)\\ &-\frac{\sqrt{2}}{3}\sigma\int_{\frac{5}{6}}^{1}\frac{1}{\sqrt{1-p}}dh(p).\end{aligned} \tag{4.16}$$

*(ii) If h is a convex function, then,*

$$\begin{aligned}\inf_{X\in V_{US}(\mu,\sigma)}\rho_h[X] =&\mu-\frac{2}{3}\sigma\int_{0}^{\frac{1}{3}}\sqrt{p}dh'(p)-\sqrt{3}\sigma\int_{\frac{1}{3}}^{\frac{2}{3}}p(1-p)dh'(p)\\ &-\frac{2}{3}\sigma\int_{\frac{2}{3}}^{1}\frac{1}{\sqrt{1-p}}dh'(p),\end{aligned} \tag{4.17}$$

*where $h'(p)$ is the right derivative of h(p).*

*(iii) If h is a general distortion function with $h(0+)=0$ and $h(1-)=1$, then,*

$$\mu-\sigma\Upsilon\left(\widetilde{(\tilde{h})^*}\right)\le\inf_{X\in V_{US}(\mu,\sigma)}\rho_h[X]\le\mu-\sigma\sup_{b\in[\frac{1}{2},1]}\Theta(h,b), \tag{4.18}$$

*where $(\tilde{h})^*$ is the concave envelope of $\tilde{h}$.*

***Proof*** See Appendix A2. □

## 5 Examples

In what follows, we discuss several important examples of distortion risk measures to illustrate the results established as preceding section. The obtained results are consistent with existing ones.

**Example 5.1** *(Case of general distributions). Letting*

$$h(p)=\min\left\{\frac{p+\beta-1}{\beta-\alpha},1\right\}1\{p\ge 1-\beta\}$$

*with $0<\alpha<\beta<1$ and $p\in[0,1]$, we get*

$$\tilde{h}(p) = 1 - h(1-p) = \begin{cases} 0, & \text{if } p \in [0, \alpha), \\ \frac{p-\alpha}{\beta-\alpha}, & \text{if } p \in [\alpha, \beta), \\ 1, & \text{if } p \in [\beta, 1], \end{cases}$$

*and*

$$\tilde{h}_*(p) = \begin{cases} 0, & \text{if } p \in [0, \alpha), \\ \frac{p-\alpha}{1-\alpha}, & \text{if } p \in [\alpha, 1). \end{cases}$$

*The graphs of the functions $\tilde{h}(p)$ and $\tilde{h}_*(p)$ are shown in Fig. 1 (left). Upon using Proposition 4.1(i), we get*

$$\sup_{X \in V(\mu,\sigma)} \text{RVaR}_{\alpha,\beta}[X] = \mu + \sigma\sqrt{\frac{\alpha}{1-\alpha}}, \tag{5.1}$$

*and the supremum in (5.1) is attained by the worst-case distribution of $X_*$ with*

$$F_{X_*}^{-1+}(p) = \begin{cases} \mu - \sigma\sqrt{\frac{1-\alpha}{\alpha}}, & \text{if } p \in [0, \alpha), \\ \mu + \sigma\sqrt{\frac{\alpha}{1-\alpha}}, & \text{if } p \in [\alpha, 1]. \end{cases}$$

*Because*

$$\text{VaR}_\alpha^+[X] = \lim_{\beta \downarrow \alpha} \text{RVaR}_{\alpha,\beta}[X], \ \text{VaR}_\alpha[X] = \lim_{\alpha' \uparrow \alpha} \text{RVaR}_{\alpha',\alpha}[X],$$

$$\text{TVaR}_\alpha[X] = \lim_{\beta \uparrow 1} \text{RVaR}_{\alpha,\beta}[X],$$

*we can directly obtain the following:*

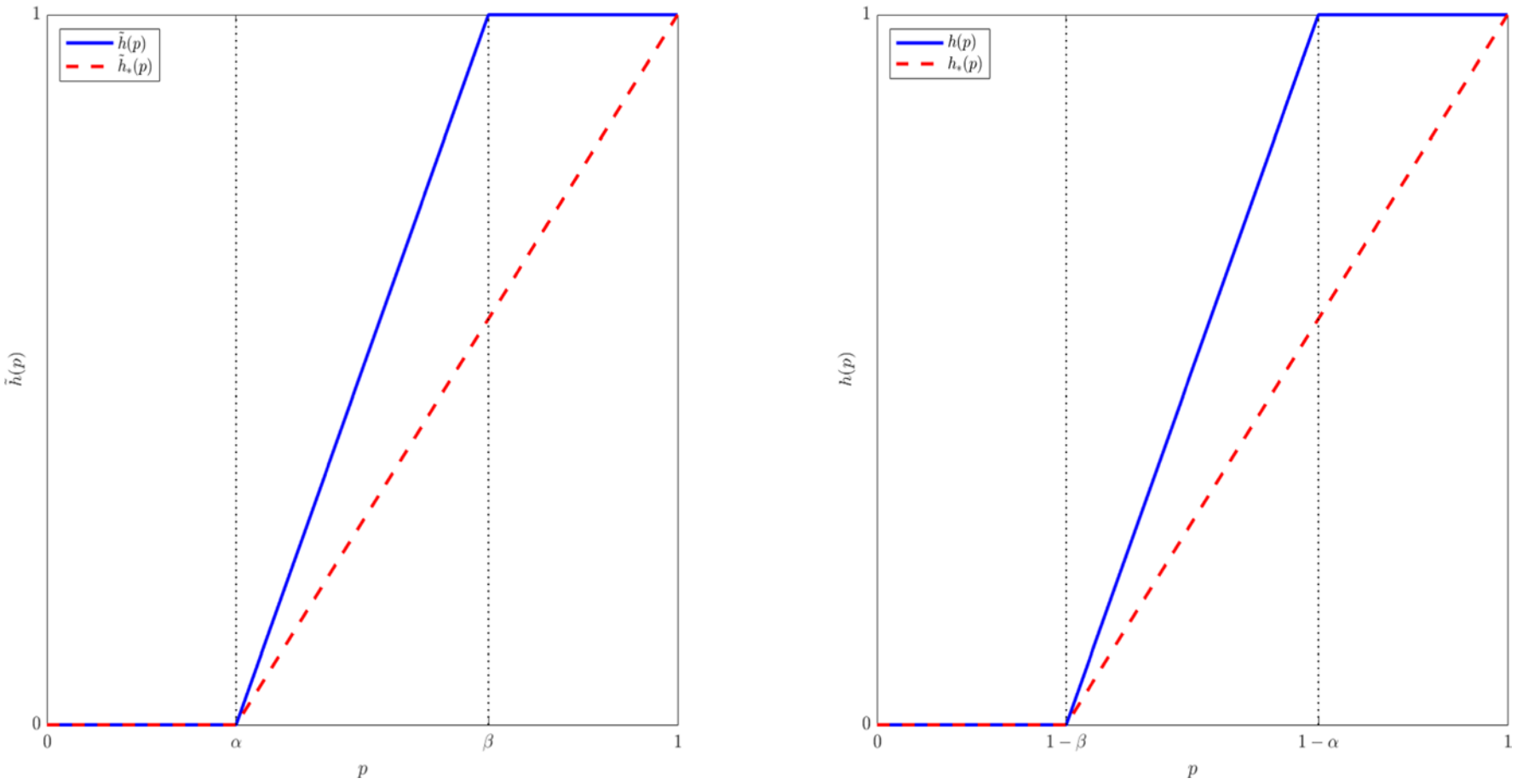

**Fig. 1** The dual distortion function and distortion function and their corresponding convex envelopes with respect to $\text{RVaR}_{\alpha,\beta}[X]$, respectively

$$\sup_{X\in V(\mu,\sigma)} \mathrm{VaR}_\alpha[X] = \sup_{X\in V(\mu,\sigma)} \mathrm{VaR}^+_\alpha[X] = \sup_{X\in V(\mu,\sigma)} \mathrm{TVaR}_\alpha[X]$$
$$= \sup_{X\in V(\mu,\sigma)} \mathrm{RVaR}_{\alpha,\beta}[X] = \mu + \sigma\sqrt{\frac{\alpha}{1-\alpha}}.$$

*While* $\sup_{X\in V(\mu,\sigma)} \mathrm{VaR}_\alpha[X]$ *cannot be attained and the last three supremums are attained by the worst-case distribution of rv* $X_*$ *with*

$$F^{-1+}_{X_*}(p) = \begin{cases} \mu - \sigma\sqrt{\frac{1-\alpha}{\alpha}}, & \text{if } p \in (0,\alpha), \\ \mu + \sigma\sqrt{\frac{\alpha}{1-\alpha}}, & \text{if } p \in [\alpha,1). \end{cases}$$

*Similarly,*

$$h(p) = \begin{cases} 0, & \text{if } p \in [0,1-\beta), \\ \frac{p+\beta-1}{\beta-\alpha}, & \text{if } p \in [1-\beta,1-\alpha), \\ 1, & \text{if } p \in [1-\alpha,1], \end{cases}$$

*and*

$$h_*(p) = \begin{cases} 0, & \text{if } p \in [0,1-\beta), \\ \frac{p+\beta-1}{\beta}, & \text{if } p \in [1-\beta,1]. \end{cases}$$

*The graphs of the functions h(p) and* $h_*(p)$ *are shown in Fig. 1 (right). Using Proposition 4.1(ii), we get*

$$\inf_{X\in V(\mu,\sigma)} \mathrm{RVaR}_{\alpha,\beta}[X] = \mu - \sigma\sqrt{\frac{1-\beta}{\beta}}, \inf_{X\in V(\mu,\sigma)} \mathrm{TVaR}_\alpha[X] = \mu,$$

$$\inf_{X\in V(\mu,\sigma)} \mathrm{VaR}_\alpha[X] = \inf_{X\in V(\mu,\sigma)} \mathrm{VaR}^+_\alpha[X] = \mu - \sigma\sqrt{\frac{1-\alpha}{\alpha}}.$$

*While* $\inf_{X\in V(\mu,\sigma)} \mathrm{VaR}^+_\alpha[X]$ *and* $\inf_{X\in V(\mu,\sigma)} \mathrm{TVaR}_\alpha[X]$ *cannot be attained; The best-case rv* $X^* = \arg\inf_{X\in V(\mu,\sigma)} \mathrm{RVaR}_{\alpha,\beta}[X]$ *is with*

$$F^{-1+}_{X^*}(p) = \begin{cases} \mu + \sigma\sqrt{\frac{\beta}{1-\beta}}, & \text{if } p \in [0,1-\beta), \\ \mu - \sigma\sqrt{\frac{1-\beta}{\beta}}, & \text{if } p \in [1-\beta,1], \end{cases}$$

*and the best-case rv* $X^* = \arg\inf_{X\in V(\mu,\sigma)} \mathrm{VaR}_\alpha[X]$ *is with*

$$F^{-1}_{X^*}(p) = \begin{cases} \mu - \sigma\sqrt{\frac{1-\alpha}{\alpha}}, & \text{if } p \in (0,\alpha], \\ \mu + \sigma\sqrt{\frac{\alpha}{1-\alpha}}, & \text{if } p \in (\alpha,1). \end{cases}$$

**Example 5.2** *(Case of symmetric distributions). We consider the distortion function* $h(p) = \min\left\{\frac{p+\beta-1}{\beta-\alpha}, 1\right\} 1\{p \geq 1-\beta\}$ *with* $0 < \alpha < \beta < 1$ *and* $p \in [0, 1]$*, the associated distortion measure given by*

$$\mathrm{RVaR}_{\alpha,\beta}[X] = \frac{1}{\beta-\alpha}\int_{\alpha}^{\beta} \mathrm{VaR}^{+}(p)dp,\ 0 < \alpha < \beta < 1.$$

*Note that*

$$\tilde{h}(p) = \min\left\{\frac{p-\alpha}{\beta-\alpha}, 1\right\} 1\{p \geq \alpha\}$$

*with* $\frac{1}{2} \leq \alpha < \beta < 1$ *and* $p \in [0, 1]$*, so we get*

$$\tilde{h}'_{*}(p) = \begin{cases} 0, & \text{if } p \in [0, \alpha), \\ \frac{1}{1-\alpha}, & \text{if } p \in [\alpha, 1), \end{cases}$$

*and*

$$\tilde{h}'_{*}(1-p) = \begin{cases} \frac{1}{1-\alpha}, & \text{if } p \in [0, 1-\alpha], \\ 0, & \text{if } p \in (1-\alpha, 1). \end{cases}$$

*By using Proposition 4.2, we get*

$$\sup_{X \in V_S(\mu,\sigma)} \mathrm{RVaR}_{\alpha,\beta}[X] = \mu + \sigma\sqrt{\frac{1}{2(1-\alpha)}}, \tag{5.2}$$

*the supremum in (5.2) is attained by the worst-case rv* $X_*$ *with*

$$F_{X_*}^{-1+}(p) = \begin{cases} \mu - \sigma\sqrt{\frac{1}{2(1-\alpha)}}, & \text{if } p \in [0, 1-\alpha], \\ \mu, & \text{if } p \in (1-\alpha, \alpha), \\ \mu + \sigma\sqrt{\frac{1}{2(1-\alpha)}}, & \text{if } p \in [\alpha, 1). \end{cases}$$

*Similarly,*

$$h'_{*}(p) = \begin{cases} 0, & \text{if } p \in [0, 1-\beta), \\ \frac{1}{\beta}, & \text{if } p \in [1-\beta, 1], \end{cases}$$

*and*

$$h'_{*}(1-p) = \begin{cases} \frac{1}{\beta}, & \text{if } p \in [0, \beta], \\ 0, & \text{if } p \in (\beta, 1]. \end{cases}$$

*By using Proposition 4.2, we get*

$$\inf_{X \in V_S(\mu,\sigma)} \mathrm{RVaR}_{\alpha,\beta}[X] = \mu - \sigma\sqrt{\frac{1-\beta}{2\beta^2}}, \tag{5.3}$$

*the best-case rv* $X^* = \arg\inf_{X \in V_S(\mu,\sigma)} \mathrm{RVaR}_{\alpha,\beta}[X]$ *is with*

$$F_{X^*}^{-1+}(p) = \begin{cases} \mu + \sigma\sqrt{\frac{1}{2(1-\beta)}}, & \text{if } p \in [0, 1-\beta), \\ \mu, & \text{if } p \in [1-\beta, \beta], \\ \mu - \sigma\sqrt{\frac{1}{2(1-\beta)}}, & \text{if } p \in (\beta, 1). \end{cases}$$

*Likewise, for* $\tilde{h}(p) = \min\left\{\frac{p-\alpha}{\beta-\alpha}, 1\right\} 1\{p \geq \alpha\}$ *with* $0 < \alpha < \beta < \frac{1}{2}$ *and* $p \in [0, 1]$*, by using Proposition 4.2, we have*

$$\sup_{X \in V_S(\mu,\sigma)} \mathrm{RVaR}_{\alpha,\beta}[X] = \mu, \tag{5.4}$$

*the supremum in (5.4) is attained by any random variable* $X_* \in V_S(\mu, \sigma)$*, and*

$$\inf_{X \in V_S(\mu,\sigma)} \mathrm{RVaR}_{\alpha,\beta}[X] = \mu - \sigma\sqrt{\frac{1}{2\beta}}, \tag{5.5}$$

*the best-case rv* $X^* = \arg\inf_{X \in V_S(\mu,\sigma)} \mathrm{RVaR}_{\alpha,\beta}[X]$ *is with*

$$F_{X^*}^{-1+}(p) = \begin{cases} \mu + \sigma\sqrt{\frac{1}{2\beta}}, & \text{if } p \in [0, 1-\beta), \\ \mu, & \text{if } p \in [1-\beta, \beta], \\ \mu - \sigma\sqrt{\frac{1}{2\beta}}, & \text{if } p \in (\beta, 1]. \end{cases}$$

**Example 5.3** *(Case of unimodal distributions). We consider the distortion function* $h(p) = \min\left\{(\frac{p}{1-\alpha})^r, 1\right\}$*, where* $0 < \alpha < 1$ *and* $0 < r \leq 1$ *are constants. Obviously, h(p) is a concave function on [0, 1]. The associated distortion measure was introduced and studied by Zhu and Li (2012).*

*(i) If* $r = 1$*, then*

$$\tilde{h}'(p) = \begin{cases} 0, & \text{if } p \in [0, \alpha), \\ \frac{1}{1-\alpha}, & \text{if } p \in [\alpha, 1). \end{cases}$$

*When* $\alpha < \frac{1}{2}$*, we have*

$$\int_0^{\frac{1}{2}} \sqrt{p(8-9p)}\, d\tilde{h}'(p) = \frac{3}{1-\alpha}\sqrt{\alpha\left(\frac{8}{9} - \alpha\right)}$$

*and*

$$\int_{\frac{1}{2}}^{1} \sqrt{(1-p)(9p-1)} d\tilde{h}'(p) = 0.$$

*When $\alpha \geq \frac{1}{2}$, we have*

$$\int_{\frac{1}{2}}^{1} \sqrt{(1-p)(9p-1)} d\tilde{h}'(p) = \frac{1}{1-\alpha}\sqrt{(1-\alpha)(9\alpha-1)}$$

*and*

$$\int_{0}^{\frac{1}{2}} \sqrt{p(8-9p)} d\tilde{h}'(p) = 0.$$

*Applying (4.8) we reobtain the following known result:*

$$\sup_{X \in V_U(\mu,\sigma)} \mathrm{TVaR}_\alpha[X] = \begin{cases} \mu + \sigma \frac{\sqrt{\alpha(8/9-\alpha)}}{1-\alpha}, & \text{if } 0 \leq \alpha < \frac{1}{2}, \\ \mu + \sigma \sqrt{\frac{8}{9(1-\alpha)} - 1}, & \text{if } \frac{1}{2} \leq \alpha < 1. \end{cases}$$

*(ii) If $0 < r < 1$, we have*

$$\tilde{h}'(p) = \begin{cases} 0, & \text{if } p \in [0,\alpha), \\ \frac{r}{1-\alpha}\left(\frac{1-p}{1-\alpha}\right)^{r-1}, & \text{if } p \in (\alpha, 1), \end{cases}$$

*and*

$$\tilde{h}''(p) = \begin{cases} 0, & \text{if } p \in [0,\alpha), \\ -\frac{r(r-1)}{(1-\alpha)^2}\left(\frac{1-p}{1-\alpha}\right)^{r-2}, & \text{if } p \in (\alpha, 1). \end{cases}$$

*Applying (4.8) we obtain the following:*
*when $\alpha < \frac{1}{2}$,*

$$\sup_{X \in V_U(\mu,\sigma)} \rho_h[X] = \mu + \frac{1}{3}\sigma \frac{r(1-r)}{(1-\alpha)^r} \int_{\alpha}^{\frac{1}{2}} \sqrt{p(8-9p)}(1-p)^{r-2} dp + \frac{1}{3}\sigma \frac{r(1-r)}{(1-\alpha)^r} \int_{\frac{1}{2}}^{1} \sqrt{9p-1}(1-p)^{r-\frac{3}{2}} dp,$$

*and, when $\alpha \geq \frac{1}{2}$,*

$$\sup_{X \in V_U(\mu,\sigma)} \rho_h[X] = \mu + \frac{1}{3}\sigma \frac{r(1-r)}{(1-\alpha)^r} \int_{\alpha}^{1} \sqrt{9p-1}(1-p)^{r-\frac{3}{2}} dp.$$

**Example 5.4** *(Case of symmetric and unimodal distributions). As in Example 5.3, we consider concave distortion function* $h(p) = \min\left\{(\frac{p}{1-\alpha})^r, 1\right\}$*, where* $0 < \alpha < 1$ *and* $0 < r \leq 1$ *are constants.*

*(1) If* $r = 1$*, then by using (4.14), we get the following known result:*

$$\sup_{X \in V_{US}(\mu,\sigma)} \mathrm{TVaR}_\alpha[X] = \begin{cases} \mu + \sigma\frac{2\sqrt{\alpha}}{3(1-\alpha)}, & \text{if } \alpha \in (0, \frac{1}{3}), \\ \mu + \sigma\sqrt{3}\alpha, & \text{if } \alpha \in [\frac{1}{3}, \frac{2}{3}), \\ \mu + \sigma\frac{2}{3}\sqrt{\frac{1}{1-\alpha}}, & \text{if } \alpha \in [\frac{2}{3}, 1). \end{cases}$$

*(2) If* $0 < r < 1$*, applying (4.14) we obtain the following:*
*when* $\alpha < \frac{1}{3}$*,*

$$\begin{aligned}
\sup_{X \in V_{US}(\mu,\sigma)} \rho_h[X] &= \mu + \sigma\frac{2r(1-r)}{3(1-\alpha)^r}\int_\alpha^{\frac{1}{3}} \sqrt{p}(1-p)^{r-2}dp \\
&\quad + \sigma\frac{\sqrt{3}r(1-r)}{(1-\alpha)^r}\int_{\frac{1}{3}}^{\frac{2}{3}} p(1-p)^{r-1}dp \\
&\quad + \sigma\frac{2r(1-r)}{3(1-\alpha)^r}\int_{\frac{2}{3}}^{1} (1-p)^{r-\frac{3}{2}}dp \\
&= \mu + \sigma\frac{2r(1-r)}{3(1-\alpha)^r}\int_\alpha^{\frac{1}{3}} \sqrt{p}(1-p)^{r-2}dp \\
&\quad + \sigma\frac{\sqrt{3}(1-r)}{(1+r)(\frac{2}{3})^r}\left((\frac{2}{3})^r(\frac{1}{3}r+1) - \frac{2r+3}{3^{1+r}}\right) \\
&\quad + \sigma\frac{4r(1-r)}{3\sqrt{3}(2r-1)};
\end{aligned}$$

*when* $\frac{1}{3} \leq \alpha < \frac{2}{3}$*,*

$$\begin{aligned}
\sup_{X \in V_{US}(\mu,\sigma)} \rho_h[X] &= \mu + \sigma\frac{\sqrt{3}r(1-r)}{(1-\alpha)^r}\int_\alpha^{\frac{2}{3}} p(1-p)^{r-1}dp \\
&\quad + \sigma\frac{2r(1-r)}{3(1-\alpha)^r}\int_{\frac{2}{3}}^{1} (1-p)^{r-\frac{3}{2}}dp \\
&= \mu + \sigma\frac{\sqrt{3}(1-r)}{(1+r)(1-\alpha)^r}\left((1-\alpha)^r(\alpha r+1) - \frac{2r+3}{3^{1+r}}\right) \\
&\quad + \sigma\frac{4r(1-r)}{3\sqrt{3}(2r-1)};
\end{aligned}$$

*and, when* $\alpha \geq \frac{2}{3}$*,*

$$\sup_{X\in V_{US}(\mu,\sigma)} \rho_h[X] = \mu + \sigma\frac{2r(1-r)}{3(1-\alpha)^r}\int_\alpha^1 (1-p)^{r-\frac{3}{2}}dp$$
$$= \mu + \sigma\frac{4r(1-r)}{3(2r-1)}(1-\alpha)^{-\frac{1}{2}}.$$

## 6 Numerical illustration

To provide a quantitative assessment of the impact of various distributional assumptions, we compute and compare the worst-case VaR and TVaR for the four uncertainty sets: $V$, $V_S$, $V_U$, and $V_{US}$. As a common parametric benchmark, we also show the corresponding values for a standard normal distribution, which satisfies all four sets and thus provides a reference point for the degree of conservatism in the worst-case bounds. To further investigate the impact of heavy-tailed risks typical in financial markets, as well as to illustrate the critical role of our finite-variance assumption, we introduce the Student's *t*-distributions with degrees of freedom $df \in \{1, 2, 5, 10\}$ as additional benchmarks. It is worth noting that the variance of a Student's *t*-distribution is given by $df/(df-2)$ for $df > 2$, which is finite, but becomes infinite for $df = 2$ and undefined for $df = 1$. Consequently, incorporating these heavy-tailed distributions allows us to examine the robustness of our bounds under standard asset returns ($df = 5, 10$), while intentionally relaxing the unit-variance constraint ($df = 1, 2$) to serve as a rigorous boundary stress-testing mechanism. Throughout these numerical evaluations, we fix the baseline constraints at $\mu = 0$ and $\sigma = 1$ without loss of generality.

Figure 2 displays the worst-case VaR and TVaR as functions of the confidence level $\alpha$ over the range [0.5, 0.999] under all four sets together with the normal and Student's *t*-distribution benchmarks. Specifically, at $\alpha = 0.99$, the worst-case VaR under $V$ reaches 9.950, which is more than four times the normal benchmark (2.326).

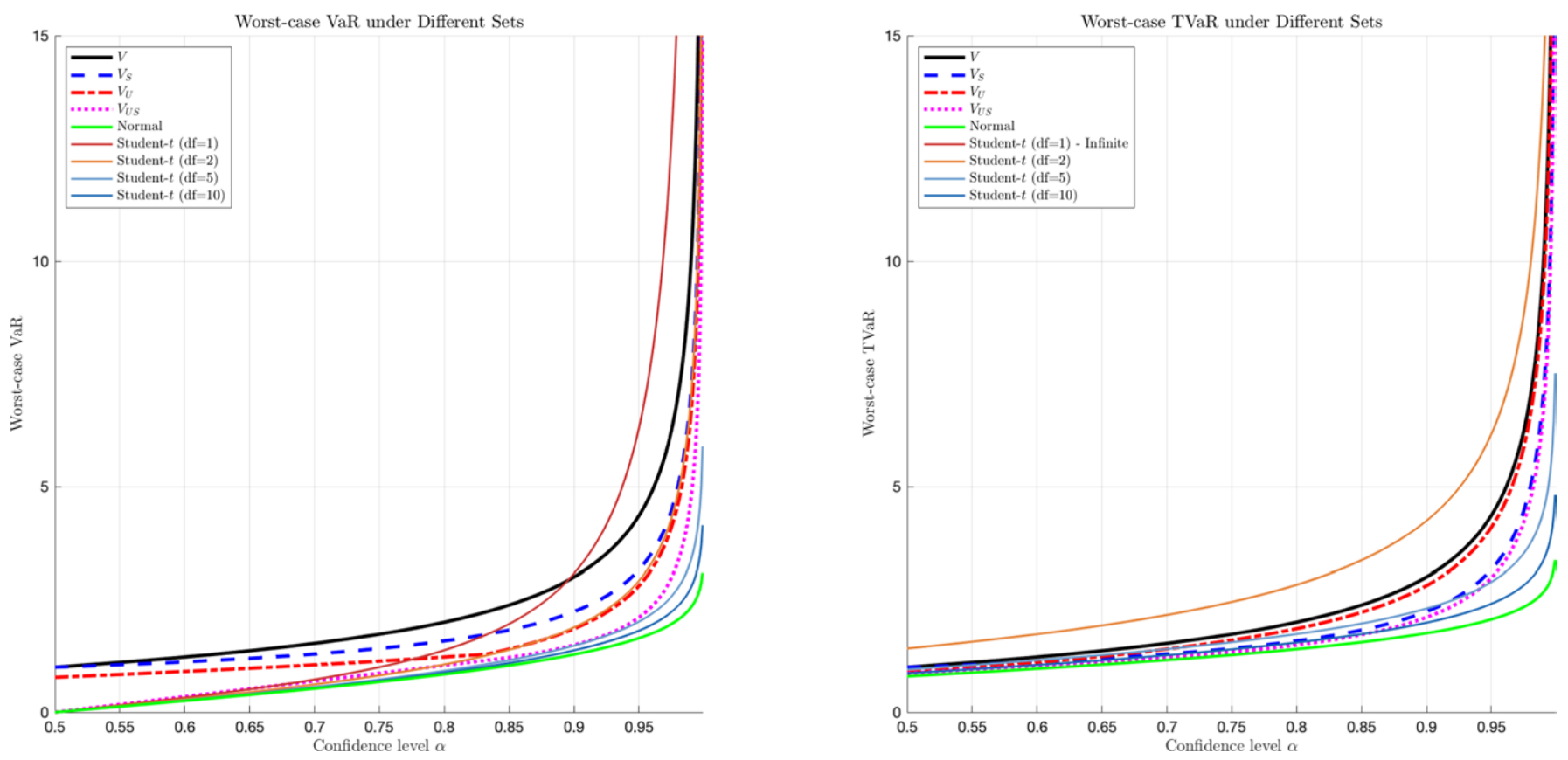

**Fig. 2** Worst-case VaR and TVaR under different distributional assumptions, compared with standard normal and Student's *t*-distributions

The inclusion of the Student's *t*-distributions provides a comprehensive illustration that highlights both the robust wrapping capacity and the critical importance of the variance constraints in risk modeling. For distributions with finite variances close to 1 (such as *t*(10) and *t*(5)), which represent heavy-tailed risks typical in financial markets, their risk curves escalate but remain safely enveloped by our robust boundaries, particularly under the symmetric-unimodal set $V_{US}$. Conversely, for $df = 1$ and $df = 2$, where the finite-variance condition is mathematically violated, the tail risks violently breach the analytical bounds at high confidence levels. This graphically demonstrates a fundamental economic intuition: while our non-parametric bounds are robustly protective against heavy-tailed innovations within the moment domain, without a bounded variance constraint, extreme tail risks become mathematically uncontrollable. Furthermore, by incorporating symmetry and unimodality ($V_{US}$), the bound drops significantly to 4.714 at $\alpha = 0.99$. This reduction of over 50% quantifies the "economic value" of shape information: while robust bounds are inherently more conservative than naive parametric estimates to account for model risk, identifying basic structural features of the loss distribution allows practitioners to substantially reduce capital buffers and improve capital efficiency.

We compare the theoretical bounds with empirical results obtained from Monte Carlo simulations. For each uncertainty set and confidence level, we performed $10^7$ independent simulations using the extremal distributions characterized in our proofs. As reported in Table 1 and Table 2, the simulated values match the analytical results with a relative deviation of less than 0.2%, confirming that the derived bounds are indeed sharp and that the extremal distributions we have identified precisely attain the theoretical limits. For the parametric benchmarks (including the standard normal distribution and the Student's *t*-distributions), their exact analytical quantiles and tail expectations are listed in the bottom rows of each table for direct comparison.

The persistent gap between our robust bounds and parametric benchmarks serves as a quantitative measure of model risk. While parametric models like the normal distribution offer simplicity (yielding a TVaR of 2.665 at $\alpha = 0.99$), they severely understate tail risk when the true underlying distribution exhibits heavy tails. For instance, the Student's *t*(5) distribution pushes the TVaR up to 4.452 at $\alpha = 0.99$, and the variance-unbounded *t*(2) distribution escalates to 14.071, completely shatter-ing the standard parametric estimates and breaching our controlled boundaries. By providing a sharp, non-parametric robust bound (e.g., 6.667 under $V_{US}$), our closed-form formulas offer a reliable safety buffer that accommodates severe model misspecifications and heavy-tailed innovations, effectively serving as a non-parametric stress-testing tool for regulators and practitioners to evaluate potential losses under the most adverse scenarios.

**Table 1** Comparison of theoretical bounds and simulated values for worst-case VaR under different uncertainty sets ($\mu = 0, \sigma = 1$)

| Uncertainty set | $\alpha = 0.90$ | | $\alpha = 0.95$ | | $\alpha = 0.99$ | |
|---|---|---|---|---|---|---|
| | Exact | Sim | Exact | Sim | Exact | Sim |
| $V$ | 3.000 | 3.000 | 4.359 | 4.359 | 9.950 | 9.950 |
| $V_S$ | 2.236 | 2.236 | 3.162 | 3.162 | 7.071 | 7.071 |
| $V_U$ | 1.856 | 1.855 | 2.809 | 2.807 | 6.591 | 6.597 |
| $V_{US}$ | 1.491 | 1.490 | 2.108 | 2.108 | 4.714 | 4.714 |
| Normal | 1.282 | 1.282 | 1.645 | 1.645 | 2.326 | 2.326 |
| Student-*t* ($df = 10$) | 1.372 | 1.372 | 1.812 | 1.812 | 2.764 | 2.764 |
| Student-*t* ($df = 5$) | 1.476 | 1.476 | 2.015 | 2.015 | 3.365 | 3.365 |
| Student-*t* ($df = 2$) | 1.886 | 1.886 | 2.920 | 2.920 | 6.965 | 6.965 |
| Student-*t* ($df = 1$) | 3.078 | 3.078 | 6.314 | 6.314 | 31.821 | 31.821 |

**Table 2** Comparison of theoretical bounds and simulated values for worst-case TVaR under different uncertainty sets ($\mu = 0, \sigma = 1$)

| Uncertainty set | $\alpha = 0.90$ | | $\alpha = 0.95$ | | $\alpha = 0.99$ | |
|---|---|---|---|---|---|---|
| | Exact | Sim | Exact | Sim | Exact | Sim |
| $\mathcal{V}$ | 3.000 | 3.000 | 4.359 | 4.359 | 9.950 | 9.950 |
| $V_S$ | 2.236 | 2.236 | 3.162 | 3.162 | 7.071 | 7.071 |
| $V_U$ | 2.809 | 2.807 | 4.096 | 4.099 | 9.375 | 9.348 |
| $V_{US}$ | 2.108 | 2.107 | 2.981 | 2.981 | 6.667 | 6.682 |
| Normal | 1.755 | 1.755 | 2.063 | 2.063 | 2.665 | 2.665 |
| Student-$t$ ($df = 10$) | 1.989 | 1.989 | 2.408 | 2.408 | 3.363 | 3.363 |
| Student-$t$ ($df = 5$) | 2.302 | 2.302 | 2.890 | 2.890 | 4.452 | 4.452 |
| Student-$t$ ($df = 2$) | 4.243 | 4.243 | 6.164 | 6.164 | 14.071 | 14.071 |
| Student-$t$ ($df = 1$) | $\infty$ | $\infty$ | $\infty$ | $\infty$ | $\infty$ | $\infty$ |

In summary, the numerical results confirm that our closed-form bounds are sharp and that the characterized extremal distributions indeed attain the theoretical limits. More importantly, the analysis demonstrates that while incorporating shape information reduces over-conservatism, a significant "robustness premium" remains necessary compared to naive parametric estimates. This unified framework thus enables practitioners to navigate the trade-off between model efficiency and distributional robustness, providing a more rigorous foundation for setting capital requirements under partial information.

## 7 Conclusions

In this work, we have obtained closed-form solutions for the extreme distortion risk measures, both worst-case and best-case, based on only the first two moments and shape information such as the symmetry/unimodality property of the underlying distribution. In addition, we have shown that the corresponding extreme-case distributions can be characterized by the envelopes of the distortion functions. Our results generalize several well-known extreme-case risk measures with closed-form solutions. Furthermore, numerical illustrations demonstrate their robustness against severe heavy-tailed distributions. The framework developed here naturally suggests several concrete avenues for future investigation. First, most of the established results can be extended to signed Choquet integrals (see Wang et al. (2020)), and it would be of interest to explore whether the extremal distributions we have characterized play a similar role in that more general setting. Second, our moment-based uncertainty set can be viewed as the limiting case of a Wasserstein ball with infinite radius; a natural next step is to study how the lower and upper bounds on DRMs adapt when, in addition to the moment constraint, the distributions in the uncertainty set are constrained to lie within a finite-radius $\sqrt{\varepsilon}$-Wasserstein ball of a given reference distribution $F$ (see, e.g., Bernard et al. (2024)), especially under the same shape constraints of symmetry and unimodality. Third, as suggested by the reviewers, while this paper focuses on the theoretical worst- and best-case bounds under partial distributional knowledge, a natural and critical next step is to conduct real data experiments, such as evaluating these robust risk measures on actual financial asset returns. Applying our closed-form formulas directly to real-world empirical data requires carefully estimating sample moments and statistically testing shape assumptions from observed time series. Developing the appropriate statistical inference procedures to address these estimation errors and calibrating our robust models to real market data represent an important and substantial direction for our future research. We are currently working on some of these problems and hope to report the findings in future papers.

## Appendix A1.

By using (3.7), the proof of statement (i) is similar to that of the proof of Theorem 4.1(i). The proof of statement (ii) is similar to that of the proof of Theorem 4.1(ii) and using (4.12). To prove (iii), for $\frac{1}{2} \leq b \leq 1$, we denote by $\mathcal{V}(b)$ the set of random variables $X$ with quantile functions

$$F_X^{-1}(p) = \begin{cases} a + c(p-1+b), \ p \in (0, 1-b), \\ a, \ p \in [1-b, b], \\ a + c(p-b), \ p \in (b, 1), \end{cases}$$

where

$$(a, b, c) \in \mathbb{R} \times \left[\frac{1}{2}, 1\right] \times \mathbb{R}^+.$$

By moments matching, we get

$$\mathcal{V}(b) \cap V(\mu, \sigma) = \left\{ X : F_X^{-1}(p) = \begin{cases} \mu + \sigma \frac{p-1+b}{\sqrt{2(1/3-b+b^2-b^3/3)}}, \ p \in (0, 1-b), \\ \mu, \ p \in [1-b, b], \\ \mu + \sigma \frac{p-b}{\sqrt{2(1/3-b+b^2-b^3/3)}}, \ p \in (b, 1), \end{cases} \right\}.$$

For $X \in \mathcal{V}(b) \cap V(\mu, \sigma)$, we have

$$\begin{aligned}\rho_h[X] =& \int_0^1 F_X^{-1}(p) d\tilde{h}(p)\\ =&\mu + c\int_0^{1-b} p d\tilde{h}(p) + c(b-1)\tilde{h}(1-b)\\ &+ c\int_b^1 p d\tilde{h}(p) - cb(1-\tilde{h}(b))\\ =&\mu + \sigma\Theta(\tilde{h}, b),\end{aligned}$$

where

$$c = \frac{\sigma}{\sqrt{\frac{2}{3}(1-b)^3}}.$$

Therefore, for this candidate distribution we obtain $\rho_h[X] = \mu + \sigma\,\Theta(\tilde{h}, b)$, which yields the lower bound $\sup_{X\in V_{US}(\mu,\sigma)} \rho_h[X] \geqslant \mu + \sigma \sup_{b\in[\frac{1}{2},1]} \Theta(\tilde{h}, b)$ in (4.15).

To obtain the upper bound of (4.15), it is well known that for any concave distortion function $h^*$, there exists a sequence of concave piecewise linear distortion functions $h_1^*(x) \leq h_2^*(x) \leq \cdots \leq h_n^*(x) \leq \cdots \leq h^*(x)$ such that $h^*(x) = \lim_{n\to\infty} h_n^*(x)$ . From the monotone convergence theorem we have $\lim_{n\to\infty} \rho_{h_n^*}[X] = \rho_{h^*}[X]$. It follows from the monotonicity of $\rho_{h_n^*}[X]$ and note that $h \leq h^*$, we have

$$\begin{aligned}\sup_{X\in V_{US}(\mu,\sigma)} \rho_h[X] \leq& \sup_{X\in V_{US}(\mu,\sigma)} \rho_{h^*}[X] = \sup_{X\in V_{US}(\mu,\sigma)} \lim_{n\to\infty} \rho_{h_n^*}[X]\\ =& \lim_{n\to\infty} \sup_{X\in V_{US}(\mu,\sigma)} \rho_{h_n^*}[X]\\ =& \lim_{n\to\infty}\left(\mu + \sigma\int_0^1 (1-p) \sup_{Z\in V_{US}(0,1)} \mathrm{TVaR}_p[Z] d(\widetilde{h_n^*})'(p)\right)\\ =&\mu + \sigma\int_0^1 (1-p) \sup_{Z\in V_{US}(0,1)} \mathrm{TVaR}_p[Z] d(\widetilde{h^*})'(p),\end{aligned}$$

from which and (4.12) we get the desired result. This completes the proof of Theorem 4.3.

## Appendix A2.

(i) Applying the result of Theorem 4.3(i), for a simple function *h*, we have

$$\sup_{Z\in V_{US}(0,1)} \rho_{\tilde{h}}[Z] =\sqrt{3}\int_{\frac{1}{2}}^{\frac{5}{6}} (2p-1)dh(p) + \frac{\sqrt{2}}{3}\int_{\frac{5}{6}}^{1} \frac{1}{\sqrt{1-p}}dh(p),$$

from which and the fact

$$\inf_{X\in V_{US}(\mu,\sigma)} \rho_h[X] = \mu - \sigma \sup_{Z\in V_{US}(0,1)} \rho_{\tilde{h}}[Z], \tag{7.1}$$

we immediately get (4.16).

(ii) If $h$ is a convex function, then by the result of Theorem 4.3(ii), we get

$$\sup_{Z\in V_{US}(0,1)} \rho_{\tilde{h}}[Z] =\frac{2}{3}\int_{0}^{\frac{1}{3}} \sqrt{p}dh'(p) + \sqrt{3}\int_{\frac{1}{3}}^{\frac{2}{3}} p(1-p)dh'(p) + \frac{2}{3}\int_{\frac{2}{3}}^{1} \frac{1}{\sqrt{1-p}}dh'(p),$$

from which and (7.1) we immediately get (4.17).

(iii) If $h$ is a general distortion function with $h(0+) = 0$ and $h(1-) = 1$, then, by (4.15),

$$\sup_{b\in[\frac{1}{2},1]} \Theta(h,b) \le \sup_{Z\in V_{US}(0,1)} \rho_{\tilde{h}}[Z] \le \Upsilon\left(\widetilde{(\tilde{h})^*}\right).$$

This, together with (7.1), immediately get (4.18).

**Acknowledgements** We would like to thank the two anonymous referees for insightful comments and constructive suggestions, which help us strengthen the paper significantly.

**Data Availability** No data was used for the research described in the article.

## Declarations

**Conflict of interest** There is no conflict of interest.